\documentclass[twocolumn]{aastex6}

\newcommand{\kms}{km$~$s$^{-1}$}
\newcommand{\vsini}{$v \sin i$}
\newcommand{\vhelio}{$v_{\rm helio}$}
\newcommand{\ali}{$A$(Li)}
\newcommand{\afe}{$A$(Fe)}
\newcommand{\cratio}{$^{12}$C/$^{13}$C}
\newcommand{\msun}{$M_{\sun}$}

\newcommand{\teff}{$T_{\rm eff}$}

\shorttitle{Lithium in Red Clump Stars}
\shortauthors{Carlberg et al.}

\defcitealias{2011ApJ...730L..12K}{K11}

\begin{document}

\title{Lithium Inventory of 2\msun\ Red Clump Stars in Open Clusters: \\ A Test of the Helium Flash  Mechanism}

\author{Joleen K. Carlberg\altaffilmark{1, 4}, Katia Cunha\altaffilmark{2},  Verne V.\ Smith\altaffilmark{3} }

\altaffiltext{1}{NASA Goddard Space Flight Center,
Code 667,
Greenbelt, MD 20771, USA
 joleen.k.carlberg@nasa.gov.}
\altaffiltext{2}{Observat\'orio Nacional, Rua General Jos\'e Cristino, 77, 20921-400 S\~ao Crist\'ov\~ao, Rio de Janeiro, RJ, Brazil}
\altaffiltext{3}{National Optical Astronomy Observatory, 950 North Cherry Avenue, Tucson, AZ 85719, USA}
\altaffiltext{4}{NASA Postdoctoral Program Fellow,  joleen.k.carlberg@nasa.gov}

\begin{abstract}
The temperature distribution of field Li-rich red giants suggests the presence of a population of Li-rich red clump (RC) stars. One proposed explanation for this population is that all stars with masses near 2~\msun\  experience a short-lived phase of Li-richness at the onset of core He-burning. Many of these stars have low \cratio, a signature of deep mixing that is presumably associated with the Li regeneration.  To test this purported mechanism of Li enrichment,  we measured abundances in  38 RC stars and 6 red giant branch (RGB) stars in four  open clusters selected to have RC masses near 2~\msun. 
We find six Li-rich stars (\ali$\geq$1.50~dex) of which only two  may be RC stars. None of the RC stars have Li exceeding the levels observed in the RGB stars, but given the brevity of the suggested Li-rich phase and the modest sample size,  it is probable that stars with larger Li-enrichments were missed simply by chance.
However, we find very few stars in our sample with low \cratio.   Such low \cratio, seen in many field Li-rich stars, should persist even  after  lithium has returned to normal low levels. 
Thus, if Li synthesis during the He flash occurs, it  is a rare, but potentially long-lived occurrence rather than a short-lived phase for all stars.  We estimate a conservative upper limit of the fraction of stars going through a Li-rich phase to be  $<47\%$, based on stars that have low \cratio\ for their observed \ali.
\end{abstract}


\section{Introduction}
A small fraction of red giant stars have lithium abundances (\ali) exceeding the predictions of standard evolution models  even though the majority of red giants  (RGs) exhibit \ali\ orders of magnitude below standard model predictions \citep{1989ApJS...71..293B}. The low \ali\ demonstrates that non-convective mixing processes contribute significantly to  Li depletion, which makes the Li-rich stars appear even more unusual.  The most Li-rich RGs have abundances exceeding the meteoritic abundances (\citealt{2000ApJ...542..978B}, \citealt{Kumar:2009gm}, \citealt{2015A&A...581A..94A}) and require a nucleosynthetic origin for the Li. 
\cite{Cameron:1971bz} described a pathway for Li nucleosynthesis  through the reactions $^3$He($\alpha$,$\gamma$)$^7$Be and $^7$Be($e^-$,$\nu$)$^7$Li. This process requires high temperatures ($>10^7$~K) for the first reaction to occur  and fast mixing (such as convection) to transport the by-products to a cool region of the star ($<3\times10^6$~K) for lithium to be long-lived.  These conditions are met at the base of the convection zones in luminous asymptotic giant branch (AGB) stars. However, 
understanding Li-rich stars\footnote{ Unless otherwise stated, ``Li-rich" in this paper refers to a RG whose \ali\ meets or exceeds the commonly used threshold of 1.5~dex.} found along the red giant branch (RGB) presents a problem because these stars' convection zones are too cool to synthesize $^7$Be. Thus, the $^3$He reaction must occur below the convection zone, and a fast non-convective mixing mechanism is required to connect the convection zone to the deeper layers of the star in order to explain the enriched surface abundances.  

Through a series of observations and advances in evolution modeling (e.g., \citealt{2000A&A...359..563C}, \citealt{2003ApJ...593..509D}, \citealt{eggleton08}), it has become largely accepted
that red giants evolving through the luminosity bump stage of evolution likely experience a short-lived phase of enriched Li. A number of deep mixing mechanisms for this stage of evolution have been proposed, including thermohaline mixing \citep{2007A&A...467L..15C}, magnetic buoyancy \citep{Busso:2007ei}, and the hybrid magneto-thermohaline mixing \citep{Denissenkov:2009ez}. 
However, as the number of the relatively rare Li-rich stars continues to grow, it has become clear that they are not restricted to just  the luminosity bump and AGB, but instead are found along the RGB. 
A new short-lived phase of Li-richness was  hypothesized by \citet[][hereafter K11]{2011ApJ...730L..12K}, who suggested that Li might be regenerated during the He-flash. This mechanism would account for the population of Li-rich red giants with temperatures too warm to be luminosity bump stars and that fall in a narrow luminosity range that coincides with the red clump (RC). This He-flash phenomenon would be relevant for stars in a narrow mass range of $\sim$1.5--2.25~\msun\ \citepalias{2011ApJ...730L..12K}, the upper limit being defined by the maximum mass that experiences a He flash, and the lower limit by the stars that maintain a relatively large reservoir of $^3$He.  For the upper bound, \cite{2016A&A...585A.124C} showed that the transition occurs closer to 1.8--1.9~\msun.
At least one Li-rich star is known to be in this mass range and is confirmed to be a He-burning star via asteroseismology \citep{2014ApJ...784L..16S}.

As a test of the hypothesis suggested by \citetalias{2011ApJ...730L..12K},  we observed RC stars in four open clusters with ages and metallicities that place their RC stars in the mass range specified by \citetalias{2011ApJ...730L..12K}. In three of the four clusters, we also observed RGs at other evolutionary stages to constrain the pre-He flash abundances. In addition to \ali, we also measure \cratio. Many of the Li-rich field stars in \citetalias{2011ApJ...730L..12K} have \cratio\ that is much lower than predicted from standard evolution models, presumably a consequence of the same mixing that brought the synthesized Li into the convection zone. Although newly synthesized Li can be destroyed if exposed to high enough temperatures, the altered \cratio\ will persist as evidence of the past deep mixing.
 
 The selection and observation of the stars in this study are described in Sections \ref{sec:clusters} and \ref{sec:obs}, respectively.  In Section \ref{sec:methods}, we outline the measurement of Li abundances and \cratio. In Section \ref{sec:discuss}, we present the significance of our results, and we give our conclusions in Section \ref{sec:end}.

\section{Cluster Selection}
\label{sec:clusters}
The four clusters in this study (Collinder~110, NGC~2204, NGC~2506, and NGC~6583) were selected so that the RC stars fell within the mass range of 1.5--2.25~\msun, based on the predictions in \citetalias{2011ApJ...730L..12K}.  Table~\ref{tab:clusters} lists the adopted cluster parameters (age, dereddened distance modulus, reddening, and metallicity) that were used to select the four clusters for study.  We used \cite{2012MNRAS.427..127B} isochrones to estimate the initial masses of the stars currently at the RC stage in each cluster, and our sample spans 1.6--2.2~\msun.  Because of uncertainties in the cluster parameters, some of the true masses may fall outside of the narrow mass range we wish to probe.

\begin{deluxetable}{llllll}
\tablecolumns{6}
\tablewidth{0pc}
\tabletypesize{\scriptsize}
\tablecaption{Literature Cluster Parameters \label{tab:clusters}}
\tablehead{
   \colhead{Cluster} &   \colhead{Age} &   \colhead{$(m-M)_0$} &   \colhead{$E(B-V)$} &   \colhead{[Fe/H]} &   \colhead{$M_{\rm RC}$} \\
    \colhead{} &   \colhead{(Gyr)  } &   \colhead{(mag)  } &    \colhead{} &   \colhead{(dex)} &       \colhead{($M_\sun$)}  } 
\startdata
Collinder 110 &  1.7\tablenotemark{c} & 11.82\tablenotemark{a} & 0.38\tablenotemark{b} &  +0.03\tablenotemark{c} & 1.9 \\
NGC 2204\tablenotemark{d}& 2.0 & 13.06 & 0.08 & $-0.23$ &  1.6  \\
NGC 2506\tablenotemark{e} & 1.99 & 12.53 & 0.04& $-0.41$ & 1.6 \\ 
NGC 6583\tablenotemark{f} & 1.0 &11.55 & 0.51\tablenotemark{g} & +0.37\tablenotemark{h} & 2.2 \\
\enddata
\tablenotetext{a}{From WEBDA and within the range of \cite{2003MNRAS.343..306B}.}
\tablenotetext{b}{\cite{2003MNRAS.343..306B}}
\tablenotetext{c}{\cite{2010AA...511A..56P}}
\tablenotetext{d}{\cite{2011AJ....141...58J}}
\tablenotetext{e}{\cite{2007AA...470..919M}}
\tablenotetext{f}{\cite{2005MNRAS.356..647C}}
\tablenotetext{g}{from $V-I$=0.63, using $E(V-I)/E(B-V)$=1.244}
\tablenotetext{h}{\cite{2010AA...523A..11M}}
\end{deluxetable}

\begin{figure*}[tb]
\includegraphics[width=0.5\textwidth]{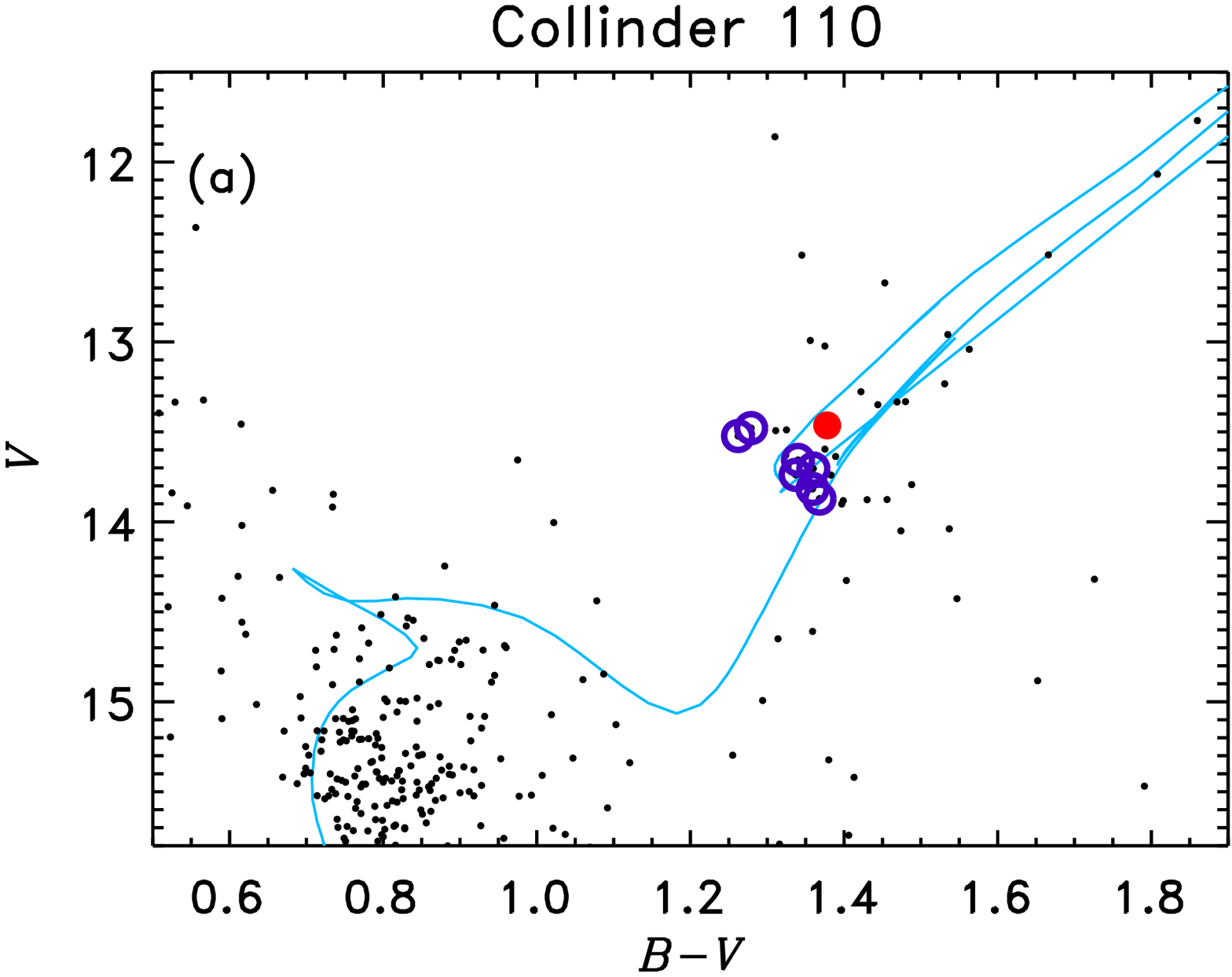}
\includegraphics[width=0.5\textwidth]{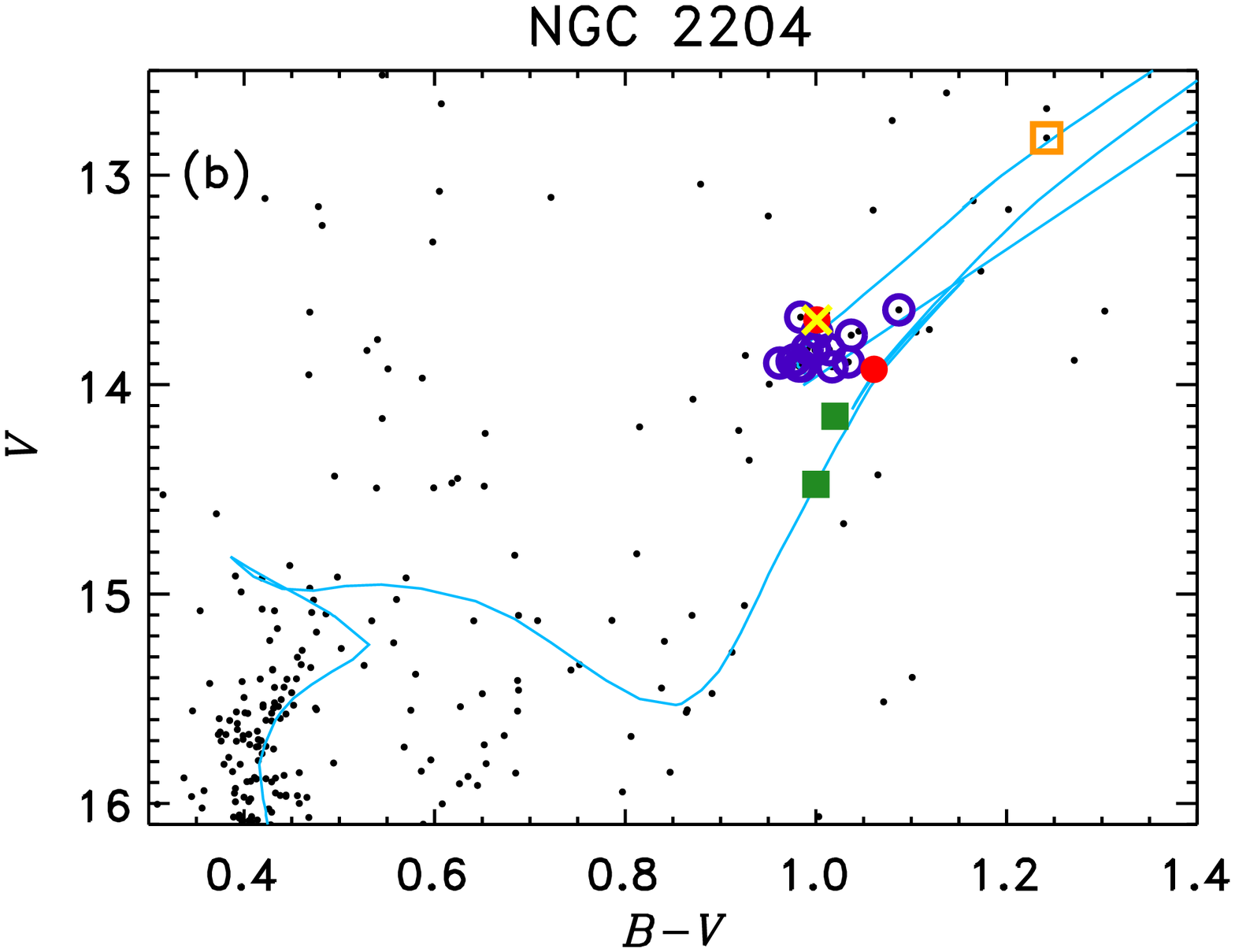}\\
\includegraphics[width=0.5\textwidth]{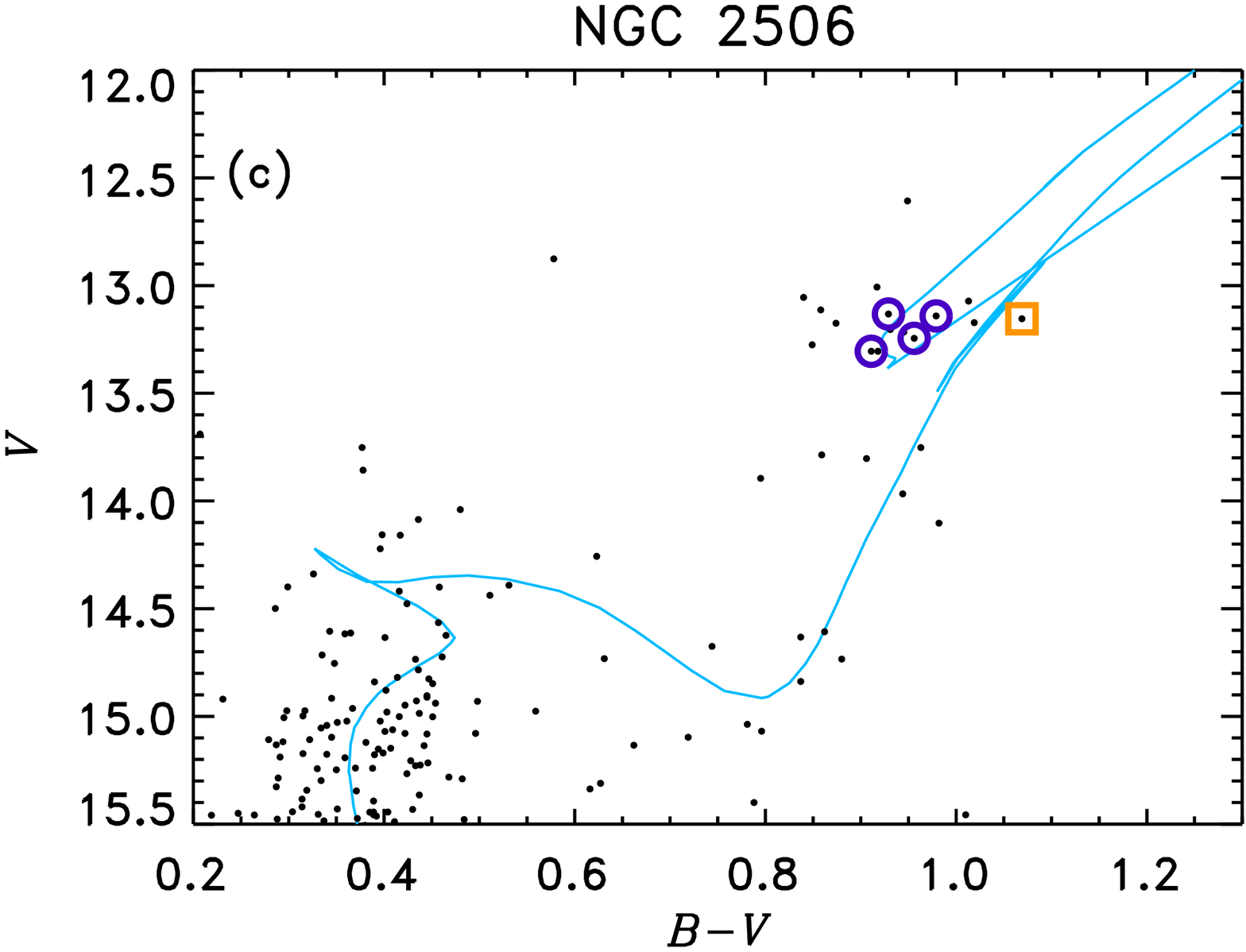}
\includegraphics[width=0.5\textwidth]{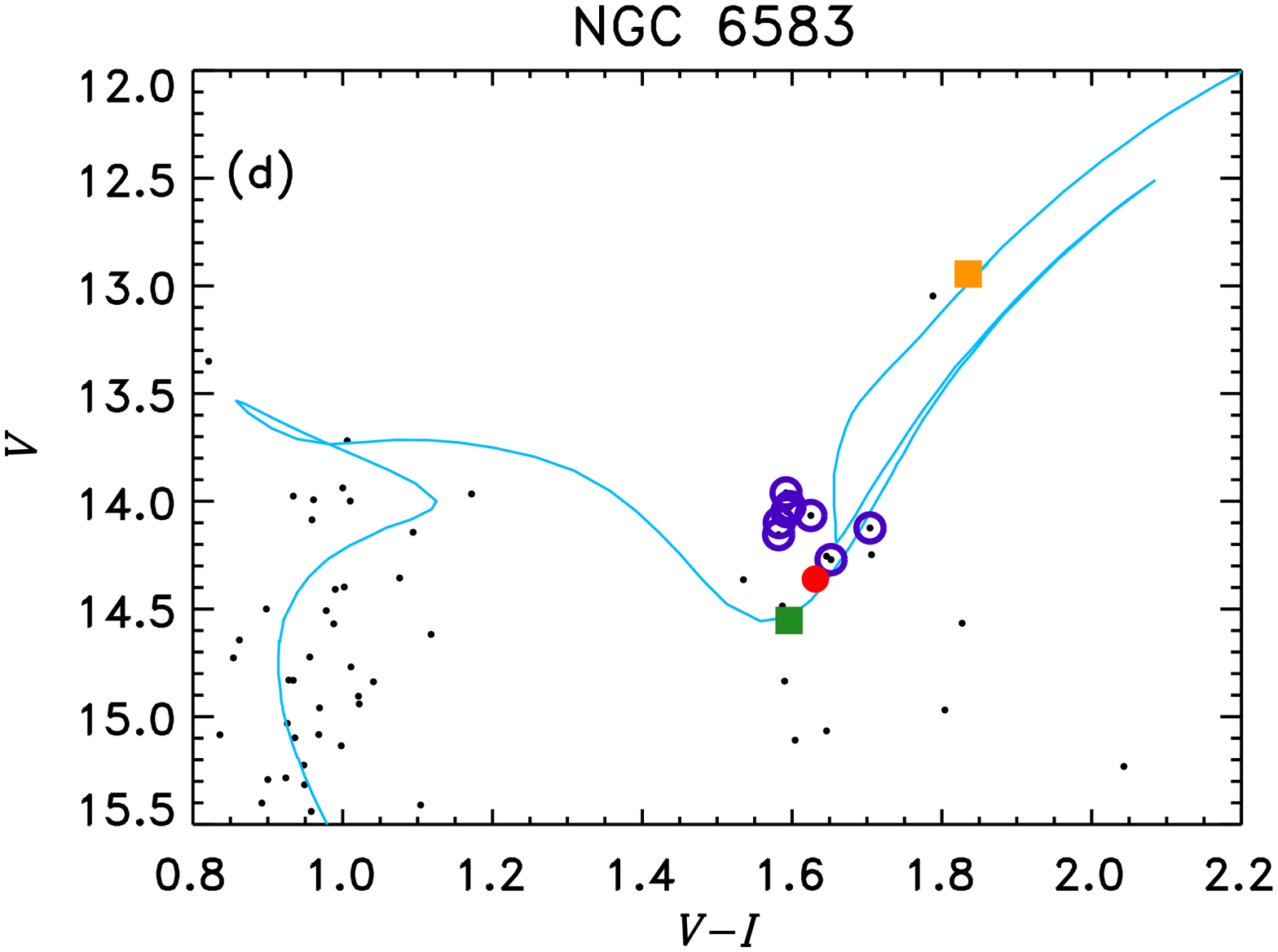}
\caption{ Color--magnitude diagrams of the four open clusters in this study. The isochrones are from \cite{2012MNRAS.427..127B} and adopt the  properties given in Table~\ref{tab:clusters} except for cluster metallicity, which comes from our calculated values in Section \ref{sec:spec_params}.
The black dots represent all candidate cluster stars, whereas large symbols denote stars analyzed in this study. RC candidate stars are circles, with the 
red, filled circles indicating  stars with high Li   (all stars with \ali$>1.5$~dex and two stars in NGC~2204 that are just below this cut-off but have have \ali\ higher than most of the other stars in the cluster) and the purple, open circles indicating stars with low Li. 
Squares represent stars selected to be comparison stars: green are lower RGB stars and orange are upper RGB stars. Filled squares have high Li, while open squares have low Li.  The yellow $\times$ in NGC 2204 is a known single-lined spectroscopic binary.
\label{fig:CMDs}}
\end{figure*}

Individual stars for each cluster were selected using color--magnitude diagrams (CMDs), as illustrated in Figure~\ref{fig:CMDs}.  The photometric data were downloaded from the WEBDA database\footnote{\url{https://www.univie.ac.at/webda/}}, and the photometric sources for each cluster can be found in Tables \ref{tab:oc_starlist1}--\ref{tab:oc_starlist4}.  The black dots represent all of the stars in the respective photometric sources, and the  large symbols represent stars observed for this study. Candidate RC stars are identified by circles, and RGs at other evolutionary stages are represented by squares.  (The meaning for the different symbol colors/shading will be described later.) Cluster isochrones \citep{2012MNRAS.427..127B} adopting ages from Table~\ref{tab:clusters} and metallicities derived in Section \ref{sec:spec_params} are also plotted.
The age of NGC~6583 is such that the red giants are at the upper edge of the critical mass range defined by \citetalias{2011ApJ...730L..12K} and higher than the upper bound suggested by \cite{2016A&A...585A.124C}, which accounts for the lack of the luminosity bump feature in the isochrone. 

\section{Observations}
\label{sec:obs}
All of the observations for this project were carried out with the Magellan Inamori Kyocera Echelle (MIKE) spectrograph on the Clay 6.5~m telescope at Las Campanas Observatory.
Depending on the night's seeing conditions and other projects being observed, we used either the 0.5\arcsec\ slit or the 0.7\arcsec\ slit, which yields resolving powers of $\sim$44,000 and 31,000, respectively.
On each night, we also observed 2--3 radial velocity (RV) standard stars. 
 We observed a Th--Ar spectrum with each pointing toward a field star or cluster. Additional Th--Ar spectra were taken approximately every 30 minutes during extended pointings toward a given cluster.
To reduce the data, we used  the Carnegie python pipeline \citep{2003PASP..115..688K}.\footnote{Available at \url{http://code.obs.carnegiescience.edu/mike}} The pipeline first performs standard CCD processing tasks, such as overscan
subtraction, bad pixel masking, flat fielding, and sky subtraction on the stellar spectra and calibration spectra. 
It extracts  both a sky spectrum and stellar spectrum from each target star observation and wavelength calibrates  these  spectra with the Th--Ar lamp spectra.  All of the echelle orders from the red arm of the spectrograph,  spanning 4800--9400~\AA,  were combined to a one-dimensional format. After measuring the RV (Section \ref{sec:rv_helio}), the spectra were velocity shifted to the stellar rest frame.  

 Tables \ref{tab:oc_starlist1}--\ref{tab:oc_starlist4} list the red giants observed in each cluster. 
 Nine stars were observed in Collinder 110 (Table~\ref{tab:oc_starlist1}) using the 0.5\arcsec\ slit, and all of these stars are RC candidates.
Nineteen stars were observed in NGC~2204 (Table~\ref{tab:oc_starlist2}) using the 0.7\arcsec\ slit, of which sixteen are RC candidate stars, two are lower RGB stars, and one is near the RGB tip.  One of the stars, MMU 4119, is a suspected spectroscopic binary \citep{2007AA...470..919M}.
In NGC~2506 (Table~\ref{tab:oc_starlist3},) only five stars were observed, of which four are RC candidates.  The fifth star is near the luminosity bump of the cluster.  All stars were observed with the 0.5\arcsec\ slit.  Finally, eleven stars were observed in NGC~6583 (Table~\ref{tab:oc_starlist4}). Both slit sizes were utilized because seeing conditions changed dramatically over the course of the night.  Nine of the stars are RC candidates, one star is at the base of the RGB, and one star is on either the upper RGB or AGB.

\begin{deluxetable*}{lllrrrrrrc}
\tablecolumns{10}
\tablewidth{0pc}
\tabletypesize{\scriptsize}
\tablecaption{Observed Red Giant Candidates: Collinder 110 \label{tab:oc_starlist1}}
\tablehead{   \colhead{Star} &  \colhead{R.A.} &   \colhead{Decl.} &   \colhead{$V$\tablenotemark{a}} &   \colhead{$B-V$\tablenotemark{a}} &
   \colhead{UT Date} &   \colhead{$t_{\rm exp}$} &   \colhead{S/N} &   \colhead{Slit}&   \colhead{Classification} \\
    \colhead{} &   \colhead{(J2000.0)} &   \colhead{(J2000.0)} &   \colhead{(mag)  } &   \colhead{(mag)  } &
    \colhead{} &   \colhead{(s)} &   \colhead{(6706~\AA)} &     \colhead{($\arcsec$)} & \colhead{(Initial/Revised)} } 
\startdata
1134   &    06:38:45.0  &     02:04:23.5   & 13.70  & 1.36 &  2013 Jan 26   &    3600  &  113 &0.5  & RC\\
2119   &    06:38:43.3  &     02:02:18.2   & 13.52 & 1.26 & 2013 Jan 27   &    4200  &  92 & 0.5 &  RC\\
2129   &    06:38:41.1  &     02:01:05.3   & 13.66 & 1.34 & 2013 Jan 26   &    3240  &118   & 0.5 & RC\\
2223   &    06:39:03.5  &     01:59:19.8   & 13.48 & 1.28 & 2013 Jan 27   &    2820  &  92 & 0.5  & RC\\
3122   &    06:38:34.7  &     02:01:41.0   & 13.46 &  1.38 & 2014 Mar 19   &    4050  &  105 & 0.5 & RC/RGB\\ 
3144   &    06:38:30.3  &     02:03:03.0   & 13.49 &  1.31 & 2014 Mar 20   &    2000  & 78 & 0.5  &  RC\\
3244   &    06:38:16.0  &     02:02:24.3   & 13.74 & 1.34 & 2013 Jan 26   &    2640  & 104  & 0.5 &  RC\\
4260   &    06:38:32.3  &     02:07:24.7   & 13.87 & 1.37 & 2013 Jan 27   &    2400  &  77 & 0.5 &  RC\\
5125   &    06:38:40.2  &     02:01:38.5   & 13.82 & 1.36 & 2013 Jan 27   &    3200  & 83  & 0.5 & RC \\
\enddata
\tablenotetext{a}{\cite{2003MNRAS.343..306B}}
\end{deluxetable*}

\begin{deluxetable*}{lllrrrrrrc}
\tablecolumns{10}
\tablewidth{0pc}
\tabletypesize{\scriptsize}
\tablecaption{Observed Red Giant Candidates: NGC~2204 \label{tab:oc_starlist2}}
\tablehead{
   \colhead{Star} &   \colhead{R.A.} &   \colhead{Decl.} &   \colhead{$V$\tablenotemark{a}} &   \colhead{$B-V$\tablenotemark{a}} &
   \colhead{UT Date} &   \colhead{$t_{\rm exp}$} &      \colhead{S/N} &   \colhead{Slit} &   \colhead{Classification} \\
    \colhead{} &   \colhead{(J2000.0)} &   \colhead{(J2000.0)} &   \colhead{(mag)  } &   \colhead{(mag)  } &    \colhead{} &   
    \colhead{(s)} &   \colhead{(6706~\AA)} &       \colhead{($\arcsec$)}   & \colhead{(Initial/Revised)}} 
\startdata
1124  &    06:15:29.0  &  $-$18:39:09.8   & 13.84 & 1.00 & 2014 Jan 06   &    1800  &  94 &0.7  & RC \\
1212  &    06:15:20.1  &  $-$18:37:57.8   & 13.88 & 0.98 & 2014 Jan 07   &    1800  &  121 &0.7  &  RC \\
1330  &    06:15:26.7  &  $-$18:33:25.1   & 13.76 & 1.04 & 2014 Jan 07   &    1400  &  113 &0.7 &  RC  \\
2212  &    06:15:49.7  &  $-$18:37:39.4   & 12.82 & 1.24 & 2014 Jan 06   &    800   & 122  &0.7 &   AGB \\
2229  &    06:15:36.9  &  $-$18:36:08.9   & 13.83 & 1.01 & 2014 Jan 07   &    1400  & 106 & 0.7 &   RC \\
2311  &    06:16:02.1  &  $-$18:38:46.5   & 13.64 & 1.09 & 2014 Jan 07   &    1200  & 109 & 0.7 &   RC \\
2330  &    06:15:34.4  &  $-$18:35:03.1   & 14.48 & 1.00 & 2014 Jan 07   &    2600  & 114 & 0.7  &  RGB \\
3205  &    06:15:46.0  &  $-$18:40:44.0   & 13.91 & 0.98 & 2014 Jan 07   &    1400  &  109 & 0.7  &  RC \\ 
3215  &    06:15:45.3  &  $-$18:43:34.9   & 13.75 & 1.00 & 2014 Jan 06   &    1400  & 114 & 0.7  &  RC \\
3321  &    06:15:43.3  &  $-$18:46:20.1   & 13.83 & 0.99 & 2014 Jan 06   &    2000  & 116 & 0.7  &  RC \\
4115  &    06:15:27.9  &  $-$18:40:34.0   & 14.15 & 1.02 & 2014 Jan 07   &    1680  & 108 & 0.7  &  RGB \\
4116  &    06:15:27.5  &  $-$18:40:13.9   & 13.93 & 1.06 & 2014 Jan 06   &    1600  & 117 & 0.7  &  RC/RGB \\
4119  &    06:15:27.2  &  $-$18:40:44.3   & 13.69 & 1.00 & 2014 Jan 07   &    1200  & 104  & 0.7  &  RC/RGB\tablenotemark{b} \\
4211  &    06:15:13.6  &  $-$18:41:49.3   & 13.68 & 0.98 & 2014 Jan 06   &    1200  & 103 & 0.7  &  RC \\
4223  &    06:15:32.9  &  $-$18:43:11.9   & 13.89  & 1.03 & 2014 Jan 06   &    1840  & 113 &  0.7  &  RC \\
4303  &    06:15:03.9  &  $-$18:41:07.6   & 13.90  & 0.96 & 2014 Jan 06   &    1400  &  100 & 0.7  &  RC \\
5352  &    06:15:50.8  &  $-$18:34:19.4   & 13.91  & 1.02 & 2014 Jan 06   &    1800  & 102 & 0.7  &  RC \\
5980  &    06:15:33.7  &  $-$18:42:12.1   & 13.89  & 0.98 & 2014 Jan 06   &    1800  & 112 & 0.7  &  RC \\
6330  &    06:15:26.1  &  $-$18:31:48.8   &  13.91 & 0.98 & 2014 Jan 06   &    1800  & 114  & 0.7 & RC \\ 
\enddata
\tablenotetext{a}{\cite{1997AJ....113.1723K}}
\tablenotetext{b}{Suspected binary}
\end{deluxetable*}

\begin{deluxetable*}{lllrrrrrrc}
\newpage
\tablecolumns{10}
\tablewidth{0pc}
\tabletypesize{\scriptsize}
\tablecaption{Observed Red Giant Candidates: NGC~2506 \label{tab:oc_starlist3}}
\tablehead{
   \colhead{Star} &   \colhead{R.A.} &   \colhead{Decl.} &   \colhead{$V$\tablenotemark{a}} &   \colhead{$B-V$\tablenotemark{a}} &   
   \colhead{UT Date} &   \colhead{$t_{\rm exp}$} &      \colhead{S/N} &\colhead{Slit}&   \colhead{Classification} \\
    \colhead{} &   \colhead{(J2000.0)} &   \colhead{(J2000.0)} &   \colhead{(mag)  } &   \colhead{(mag)  } &    \colhead{} &
   \colhead{(s)} &   \colhead{(6706~\AA)} &       \colhead{($\arcsec$)}  & \colhead{(Initial/Revised)} } 
\startdata
2380  &    08:00:09.2  &  $-$10:49:09.0   & 13.14 & 0.98 &  2013 Jan 27   &    1680  & 68 & 0.5  &  RC\\
3265  &    07:59:50.8  &  $-$10:46:40.0   & 13.15 & 1.07 & 2013 Jan 27   &    1320  &  52 & 0.5  & RGB/RC?\tablenotemark{b} \\
4138  &    08:00:01.4  &  $-$10:45:39.3   & 13.31 & 0.91 & 2013 Jan 27   &    2220  & 78  & 0.5  &  RC\\
4205  &    07:59:51.3  &  $-$10:46:17.4   & 13.24 & 0.96 & 2013 Jan 26   &    1680  &  96  & 0.5  &  RC\\
4240  &    07:59:52.6  &  $-$10:44:50.0   & 13.13 & 0.93 & 2013 Jan 26   &    1620  &  94 &  0.5  &  RC\\
\enddata
\tablenotetext{a}{\cite{2001AcA....51...49K}}
\tablenotetext{b}{Suspected binary}
\end{deluxetable*}

\begin{deluxetable*}{lllrrrrrrc}
\tablecolumns{10}
\tablewidth{0pc}
\tabletypesize{\scriptsize}
\tablecaption{Observed Red Giant Candidates: NGC~6583 \label{tab:oc_starlist4}}
\tablehead{
   \colhead{Star} &   \colhead{R.A.} &   \colhead{Decl.} &   \colhead{$V$\tablenotemark{a}} &   \colhead{$V-I$\tablenotemark{a}} &
   \colhead{UT Date} &   \colhead{$t_{\rm exp}$} &   \colhead{S/N} &   \colhead{Slit} &   \colhead{Classification} \\
    \colhead{} &   \colhead{(J2000.0)} &   \colhead{(J2000.0)} &   \colhead{(mag)  } &   \colhead{(mag)  } &  
    \colhead{} &   \colhead{(s)} &   \colhead{(6706~\AA)} &       \colhead{($\arcsec$)}  & \colhead{(Initial/Revised)} } 
\startdata
10    &    18:15:51.1  &  $-$22:07:15.2   & 12.94 & 1.84 & 2014 Jun 19   &    840   &  112 &0.5 &  AGB/RGB  \\
33    &    18:15:52.6  &  $-$22:09:53.2   & 14.02 & 1.60 & 2014 Jun 19   &    2280  & 109 & 0.5  &  RC \\
34    &    18:15:43.9  &  $-$22:09:00.9   & 13.96 & 1.59 & 2014 Jun 19   &    1920  & 100  &0.5  &  RC\\
38    &    18:15:48.2  &  $-$22:09:53.2   & 14.05 & 1.59 & 2014 Jun 19   &    1600  &  129 &0.7  &  RC\\
39    &    18:15:56.6  &  $-$22:07:32.4   & 14.07 & 1.62 & 2014 Jun 19   &    1800  &  119 &0.7 &   RC\\
42    &    18:15:50.2  &  $-$22:09:55.5   & 14.12 & 1.70 & 2014 Jun 19   &    900   &  70 &0.7  &  RC\\
46    &    18:15:51.2  &  $-$22:07:26.6   & 14.10 & 1.58 & 2014 Jun 19   &    2400  & 96  &0.5  &  RC\\
50    &    18:15:48.3  &  $-$22:09:33.5   & 14.15 & 1.58 & 2014 Jun 19   &    1800  & 106  &0.7  &  RC/RGB?\\
62    &    18:15:51.2  &  $-$22:08:28.1   & 14.27 & 1.65 & 2014 Jun 19   &    2000  & 109  &0.7  &  RC\\
72    &    18:15:54.3  &  $-$22:08:05.7   & 14.36 & 1.63 & 2014 Jun 19   &    2200  & 98  &0.7  &  RC/RGB?\\
92    &    18:15:51.7  &  $-$22:08:30.6   & 14.55 & 1.60 & 2014 Jun 19   &    3800  & 107  &0.5 & RGB \\
\enddata
\tablenotetext{a}{\cite{2005MNRAS.356..647C}}
\end{deluxetable*}

\section{Stellar Characterization}
\label{sec:methods}

\subsection{Radial Velocities}
\label{sec:rv_helio}
Heliocentric radial velocities (\vhelio) were measured for each  star by cross-correlating the 1D spectra with the high-resolution Arcturus atlas spectrum \citep{hinkle00} with the IRAF procedure {\it fxcor}.  The spectra were also cross-correlated with the \cite{hinkle00} telluric spectrum in the wavelength ranges of 6475--6675\AA, 6980--7060~\AA, and 7850--8125~\AA\ to measure slit centering errors, since the seeing disk was at times smaller than the slit width. 
 We found   RV offsets in the telluric line centers as high as 2.4~\kms.  The final \vhelio\ measurements reported here are $v_{\rm helio} = v_{\rm stellar} + v_{\rm cor} - v_{\rm telluric}$, where  $v_{\rm stellar}$ is the relative RV of the object spectra with respect to the Arcturus spectrum (which is in the stellar rest frame), $v_{\rm cor}$ is the heliocentric RV correction, and $v_{\rm telluric}$ is the relative RVs of the telluric lines in the observed spectra with respect to the atlas telluric spectrum. 
 Using the corrections from the telluric line cross-correlation  better reproduced the \vhelio\  of the RV standard stars that we observed. 

The average cluster velocities were calculated using an iterative sigma-clipping algorithm, clipping at $3\sigma$. The results
are given in Table~\ref{tab:oc_rv}, along with literature measurements.  Two stars in NGC~2204 were clipped:  2330 with $v_{\rm helio}=96.5$~\kms\  and 4119 with $v_{\rm helio}=76.8$~\kms. The latter is a suspected spectroscopic binary star that is still a cluster member \citep{2007AA...470..919M}.  We maintain star 2330 in our analysis, but caution that it is a potential non-member based on its RV. If it is a member, it is on the lower RGB. Individual \vhelio\ are listed in the final column of Table \ref{tab:stell_param}. The uncertainty in the \vhelio\ is dominated by systematic errors. The random errors are all $<0.1$~\kms\ with an average 0.04~\kms. Comparing the \vhelio\ of five RV standard stars (HD~107328, HD~171391, HD~66141, HD~26162, and HD~203638) measured on multiple nights to the accepted values in ``The Astronomical Almanac'' \citep{2006asal.book.....U}, we estimate a systematic uncertainty of 0.2--0.3~\kms.

\begin{deluxetable*}{lrrr|rrrr}
\tablecolumns{4}
\tablewidth{0pc}
\tabletypesize{\scriptsize}
\tablecaption{Comparison of Measured Open Cluster Properties \label{tab:oc_rv}}
\tablehead{
   \colhead{Cluster Name} &
   \colhead{$v_{\rm helio}$} &
   \colhead{Lit. $v_{\rm helio}$} &
   \colhead{Reference} &
   \colhead{[Fe/H]} &
   \colhead{$\sigma_{\rm [Fe/H]}$} &
   \colhead{Lit. [Fe/H]} &
   \colhead{Reference}\\
      \colhead{} &
   \colhead{(\kms)} &
   \colhead{(\kms)} &
   \colhead{} &
   \colhead{(dex)} &
   \colhead{(dex)} &
   \colhead{(dex)} &
   \colhead{}  }
\startdata
Collinder 110      &$+37.0\pm1.3$   &   $+41.0\pm3.8$& 1 & $-0.09$ & 0.04 & +0.03 & 1  \\
                           &    \nodata                       &   $+38.7\pm0.8$ & 2  &  \nodata &  \nodata & \nodata & \nodata\\
NGC~2204           & $+91.1\pm1.2$  & $+88.4\pm1.3$ & 3  & $-0.21$ &0.04  & $-0.23$ & 3\\
                            & \nodata                           & $+91.4\pm1.3$ & 4  &  \nodata&  \nodata & $-0.32$ & 7 \\
NGC~2506           & $+82.4\pm1.1$  & $+83.2\pm1.6$ & 4, 5  & $-0.25$ & 0.09  & $-0.41$& 4 \\
                           &  \nodata                         &  \nodata   & \nodata  &\nodata  &  \nodata    &  $-0.44$ & 7 \\
                           &  \nodata                         &  \nodata   &  \nodata &\nodata  &  \nodata     &  $-0.24$ & 8, 9 \\
NGC~6583           & $-2.3\pm1.1$     & $-3.0\pm0.4$ & 6  & $+0.17$  & 0.05 & $+0.37$ & 6 \\
\enddata
\tablerefs{(1)  \cite{2010AA...511A..56P},  (2) \cite{2014AJ....147..138C}, (3) \cite{2011AJ....141...58J},  (4) \cite{2007AA...470..919M}, (5) \cite{2008AA...485..303M}, (6) \cite{2010AA...523A..11M}, (7) \cite{2002AJ....124.2693F}, (8) \cite{2011MNRAS.416.1092M}, (9) \cite{2012MNRAS.425.1567L}}
\end{deluxetable*}

\subsection{ Spectroscopic Stellar Parameters}
\label{sec:spec_params}
Although it is common in open cluster studies to use photometric temperatures and isochrones to constrain surface gravities, we opted to  do a full spectroscopic derivation of the stellar parameters. This  provides a more homogenous determination of stellar \teff\ and $\log g$ across our sample.
We used the iron line list of \cite{carlberg12}, which was compiled for deriving stellar parameters of  red giant stars.   Equivalent widths (EqWs) were all measured by hand using IRAF's {\it splot} routine, using gaussian profiles and de-blending when necessary. The high-resolution Arcturus atlas \citep{hinkle00} was used as a template to identify blended lines.  We used the 2014 version of the stellar line analysis program, MOOG\footnote{available at \url{http://www.as.utexas.edu/~chris/moog.html}} 
\citep{Sneden:1973el}, 
and MARCS spherical
atmosphere models  \citep{2008PhST..133a4003P} to compute abundances. To solve for the stellar $T_{\rm eff}$, $\log g$, [Fe/H], and $\xi$, we used the standard requirements that there should be no trend of the output $A$(\ion{Fe}{1}) with either the excitation potential or strength of the iron lines and that  $A$(\ion{Fe}{1})  should equal $A$(\ion{Fe}{2}).    The 2014 version of MOOG  adopts \afe$_\sun=7.50$, while the MARCS models adopt \afe$_\sun=7.45$.
However, the difference in the atmosphere structure of two models that differ in [M/H] by 0.05~dex  is too small to affect our abundances.
The \cite{carlberg12}  line list yields a slightly higher solar abundance of \afe$_\sun=7.53$, which we adopt to translate our stellar \afe\ to [Fe/H].
In Table~\ref{tab:stell_param}, we give the results of our stellar parameter analysis and the uncertainties in each parameter. 
The uncertainties in  $\xi$ are the variations that arise when the slope of $A$(\ion{Fe}{1}) vs.\ EqW/$\lambda$ is varied within the 1$\sigma$ uncertainty of the fitted slope  \citep{1997A&A...328..261N}. Similarly, the uncertainty in \teff\ uses the  uncertainty in the slope
of $A$(\ion{Fe}{1}) vs.\  excitation potential. It also includes the contribution from the $\xi$ uncertainty.
The uncertainties in [Fe/H] and $\log g$ are  the standard deviations in the 
\ion{Fe}{1} and \ion{Fe}{2} lines, respectively. These dominate over the uncertainty arising from sensitivities to the other stellar parameters.

\begin{deluxetable*}{lrrrrrrrrrrrrr}
\tablecolumns{11}
\tablewidth{0pc}
\tabletypesize{\scriptsize}
\tablecaption{Stellar Parameters \label{tab:stell_param}}
\tablehead{
   \colhead{Cluster} &
   \colhead{Star} &
   \colhead{\teff} &
   \colhead{$\sigma_{T_{\rm eff}}$} &
   \colhead{$\log g$} &
   \colhead{$\sigma_{\log g}$} &
   \colhead{[Fe/H]} &
   \colhead{$\sigma_{\rm [Fe/H]}$} &
   \colhead{$\xi$} &
   \colhead{$\sigma_\xi$} &
   \colhead{\vsini} & 
   \colhead{$\sigma_{v \sin i}$\tablenotemark{a}} &
   \colhead{$\zeta$\tablenotemark{b}} &
   \colhead{\vhelio} \\ 
   \colhead{} &
   \colhead{} &
   \colhead{(K)} &
   \colhead{(K)} &
   \colhead{(dex)} &
   \colhead{(dex)} &
   \colhead{(dex)} &
   \colhead{(dex)} &
   \colhead{(\kms)} &
   \colhead{(\kms)} &
   \colhead{(\kms)} &
   \colhead{(\kms)} &
   \colhead{(\kms)} &
   \colhead{(\kms)}  }
\startdata
Collinder 110 & 1134 & 4960. &  62. & 2.80 & 0.09 & $-0.13$ & 0.10 & 1.47 & 0.06 & 1.0 & 0.0 & 4.99 & +37.2 \\
Collinder 110 & 2119 & 4970. &  64. & 2.80 & 0.14 & $-0.16$ & 0.10 & 1.14 & 0.06 & 5.9 & 0.2 & 5.23 & +35.6 \\
Collinder 110 & 2129 & 4940. &  59. & 2.70 & 0.09 & $-0.11$ & 0.10 & 1.35 & 0.05 & 1.1 & 0.0 & 4.98 & +38.0 \\
Collinder 110 & 2223 & 4990. &  69. & 2.80 & 0.07 & $-0.07$ & 0.12 & 1.43 & 0.06 & 1.0 & 0.0 & 5.34 & +36.2 \\
Collinder 110 & 3122 & 4800. &  80. & 2.65 & 0.10 & $-0.10$ & 0.13 & 1.33 & 0.07 & 1.0 & 0.0 & 4.81 & +39.2 \\
Collinder 110 & 3144 & 4870. &  72. & 2.55 & 0.12 & $-0.07$ & 0.12 & 1.37 & 0.06 & 1.4 & 0.0 & 4.97 & +37.2 \\
Collinder 110 & 3244 & 4960. &  62. & 2.85 & 0.09 & $-0.08$ & 0.10 & 1.39 & 0.06 & 1.0 & 0.0 & 4.95 & +38.0 \\
Collinder 110 & 4260 & 4980. &  64. & 2.75 & 0.13 & $-0.10$ & 0.11 & 1.19 & 0.06 & 1.0 & 0.0 & 4.85 & +36.7 \\
Collinder 110 & 5125 & 5130. &  72. & 3.05 & 0.11 & +0.00 & 0.12 & 1.45 & 0.07 & 1.0 & 0.0 & 5.39 & +34.6 \\
NGC 2204 & 1124 & 5030. &  56. & 2.65 & 0.13 & $-0.24$ & 0.10 & 1.44 & 0.06 & 2.7 & 0.0 & 5.43 & +91.2 \\
NGC 2204 & 1212 & 5080. &  55. & 2.75 & 0.08 & $-0.21$ & 0.10 & 1.28 & 0.05 & 2.5 & 0.0 & 5.54 & +88.1 \\
NGC 2204 & 1330 & 5120. &  71. & 2.85 & 0.10 & $-0.18$ & 0.12 & 1.43 & 0.07 & 1.0 & 0.0 & 5.83 & +88.9 \\
NGC 2204 & 2212 & 4570. &  62. & 2.00 & 0.10 & $-0.27$ & 0.10 & 1.55 & 0.05 & 1.9 & 0.0 & 5.21 & +91.7 \\
NGC 2204 & 2229 & 5060. &  52. & 2.70 & 0.14 & $-0.18$ & 0.09 & 1.35 & 0.05 & 1.5 & 0.0 & 5.54 & +91.3 \\
NGC 2204 & 2311 & 4950. &  61. & 2.55 & 0.13 & $-0.26$ & 0.11 & 1.46 & 0.06 & 1.5 & 0.0 & 5.40 & +91.8 \\
NGC 2204 & 2330 & 5080. &  59. & 3.15 & 0.10 & $-0.17$ & 0.10 & 1.23 & 0.06 & 1.0 & 0.0 & 4.83 & +96.5 \\
NGC 2204 & 3205 & 5070. &  72. & 2.85 & 0.07 & $-0.21$ & 0.12 & 1.38 & 0.07 & 1.0 & 0.0 & 5.47 & +93.3 \\
NGC 2204 & 3215 & 5010. &  61. & 2.75 & 0.06 & $-0.17$ & 0.10 & 1.42 & 0.06 & 1.0 & 0.0 & 5.45 & +91.5 \\
NGC 2204 & 3321 & 5100. &  58. & 2.80 & 0.11 & $-0.15$ & 0.10 & 1.32 & 0.06 & 2.3 & 0.0 & 5.67 & +91.3 \\
NGC 2204 & 4115 & 4880. &  63. & 2.70 & 0.09 & $-0.21$ & 0.10 & 1.28 & 0.06 & 2.2 & 0.0 & 4.62 & +91.5 \\
NGC 2204 & 4116 & 4860. &  63. & 2.65 & 0.11 & $-0.26$ & 0.11 & 1.35 & 0.06 & 1.0 & 0.0 & 4.80 & +92.9 \\
NGC 2204 & 4119 & 4820. &  71. & 2.65 & 0.14 & $-0.24$ & 0.11 & 1.27 & 0.06 & 1.0 & 0.0 & 4.96 & +76.8 \\
NGC 2204 & 4211 & 5030. &  65. & 2.70 & 0.10 & $-0.19$ & 0.11 & 1.41 & 0.06 & 1.0 & 0.0 & 5.63 & +90.7 \\
NGC 2204 & 4223 & 5260. & 101. & 3.06 & 0.11 & $-0.15$ & 0.17 & 1.62 & 0.10 & 1.3 & 0.0 & 6.15 & +91.0 \\
NGC 2204 & 4303 & 5010. &  71. & 2.60 & 0.10 & $-0.26$ & 0.12 & 1.47 & 0.07 & 1.0 & 0.0 & 5.27 & +91.3 \\
NGC 2204 & 5352 & 5010. &  68. & 2.80 & 0.09 & $-0.15$ & 0.11 & 1.36 & 0.06 & 1.0 & 0.0 & 5.25 & +90.6 \\
NGC 2204 & 5980 & 5050. &  63. & 2.70 & 0.10 & $-0.23$ & 0.11 & 1.46 & 0.06 & 1.8 & 0.0 & 5.43 & +91.0 \\
NGC 2204 & 6330 & 5000. &  63. & 2.70 & 0.06 & $-0.25$ & 0.11 & 1.42 & 0.06 & 1.0 & 0.0 & 5.22 & +90.9 \\
NGC 2506 & 2380 & 4980. &  65. & 2.55 & 0.09 & $-0.32$ & 0.11 & 1.45 & 0.06 & 1.0 & 0.0 & 5.31 & +82.8 \\
NGC 2506 & 3265 & 5250. & 100. & 3.05 & 0.15 & $-0.10$ & 0.17 & 1.52 & 0.10 & 1.0 & 0.0 & 6.24 & +80.3 \\
NGC 2506 & 4138 & 5030. &  64. & 2.70 & 0.09 & $-0.29$ & 0.11 & 1.45 & 0.06 & 1.0 & 0.0 & 5.27 & +82.9 \\
NGC 2506 & 4205 & 5050. &  61. & 2.80 & 0.08 & $-0.25$ & 0.10 & 1.41 & 0.06 & 1.0 & 0.0 & 5.42 & +83.7 \\
NGC 2506 & 4240 & 5000. &  66. & 2.70 & 0.09 & $-0.27$ & 0.11 & 1.43 & 0.06 & 1.0 & 0.0 & 5.38 & +82.4 \\
NGC 6583 &   10 & 4350. & 103. & 2.10 & 0.28 & +0.05 & 0.15 & 1.61 & 0.09 & 2.3 & 0.6 & 4.26 & $0.0$ \\
NGC 6583 &   33 & 4870. &  81. & 2.80 & 0.12 & +0.19 & 0.13 & 1.60 & 0.07 & 1.5 & 0.0 & 4.54 & $-3.2$ \\
NGC 6583 &   34 & 4860. &  73. & 2.65 & 0.11 & +0.16 & 0.12 & 1.60 & 0.06 & 1.9 & 0.0 & 4.58 & $-2.7$ \\
NGC 6583 &   38 & 4900. &  75. & 2.75 & 0.16 & +0.18 & 0.12 & 1.54 & 0.07 & 2.1 & 0.0 & 4.59 & $-1.7$ \\
NGC 6583 &   39 & 4830. &  76. & 2.50 & 0.12 & +0.15 & 0.12 & 1.64 & 0.06 & 5.0 & 0.0 & 4.38 & $-1.0$ \\
NGC 6583 &   42 & 4840. &  90. & 2.60 & 0.23 & +0.18 & 0.14 & 1.68 & 0.08 & 3.3 & 0.0 & 4.36 & $-2.5$ \\
NGC 6583 &   46 & 4910. &  77. & 2.80 & 0.13 & +0.17 & 0.12 & 1.56 & 0.07 & 1.5 & 0.0 & 4.56 & $-2.5$ \\
NGC 6583 &   50 & 4970. &  85. & 3.00 & 0.16 & +0.27 & 0.13 & 1.64 & 0.08 & 2.5 & 0.0 & 4.67 & $-2.8$ \\
NGC 6583 &   62 & 4910. &  84. & 2.80 & 0.15 & +0.20 & 0.13 & 1.61 & 0.07 & 2.0 & 0.0 & 4.39 & $-1.8$ \\
NGC 6583 &   72 & 4940. &  92. & 2.95 & 0.16 & +0.16 & 0.14 & 1.55 & 0.08 & 5.0 & 0.2 & 4.37 & $-2.7$ \\
NGC 6583 &   92 & 4980. &  71. & 2.95 & 0.14 & +0.19 & 0.11 & 1.42 & 0.06 & 1.1 & 0.0 & 4.28 & $-4.1$ \\
\enddata
\tablenotetext{a}{Values of 0.0 refer to upper limits.}
\tablenotetext{b}{Adopted}
\end{deluxetable*}

\begin{figure*}
\includegraphics[width=0.5\textwidth]{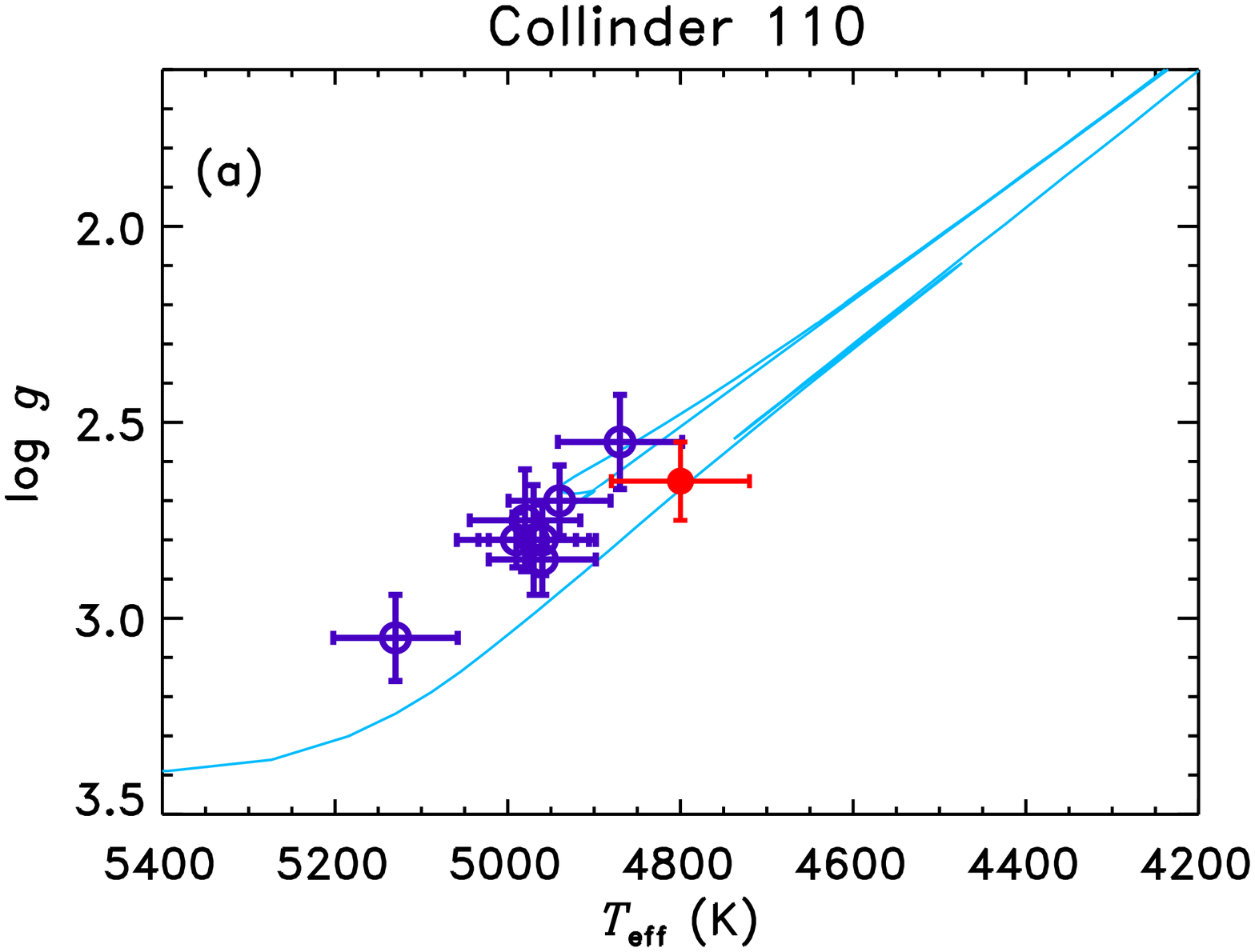}\includegraphics[width=0.5\textwidth]{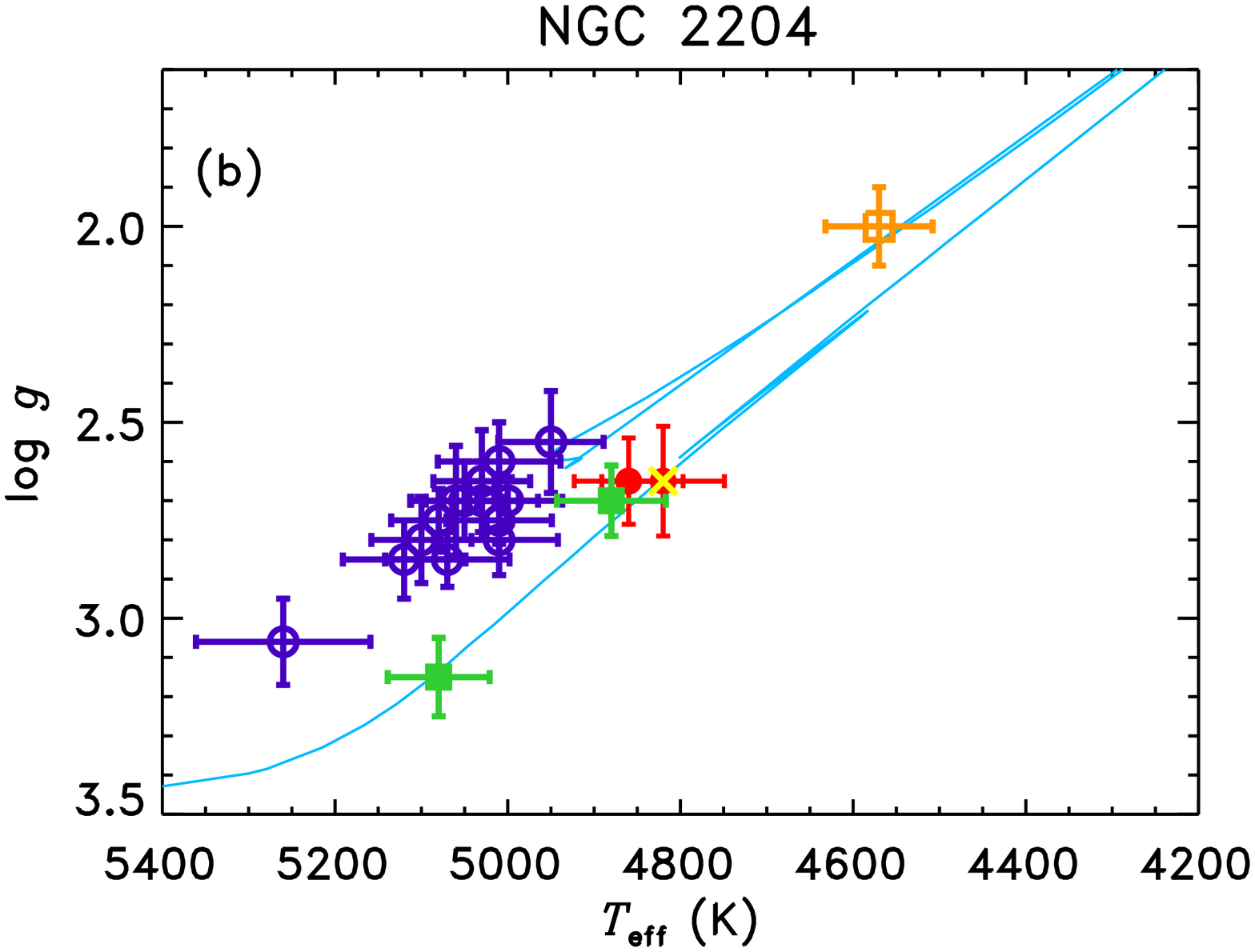}\\
\includegraphics[width=0.5\textwidth]{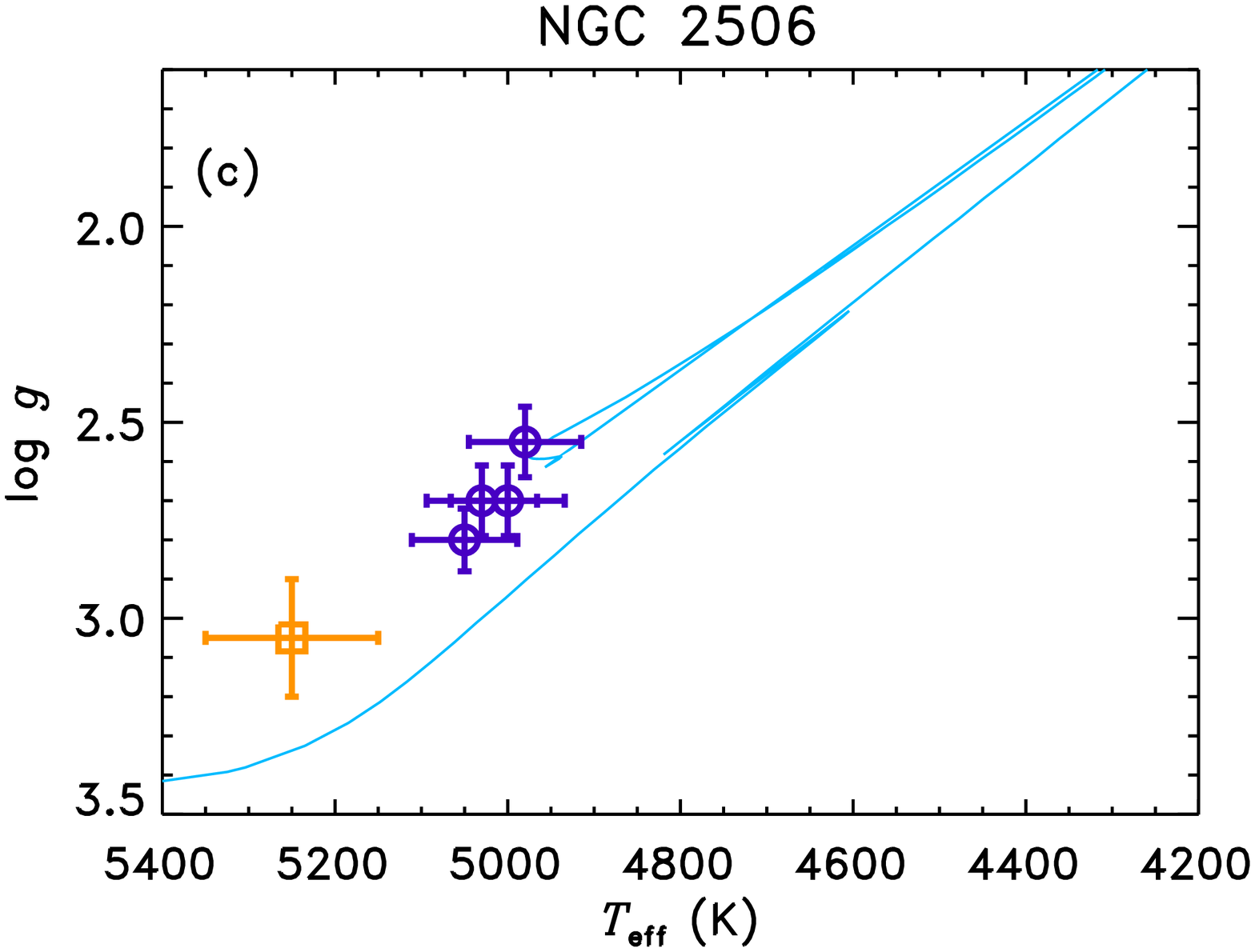}\includegraphics[width=0.5\textwidth]{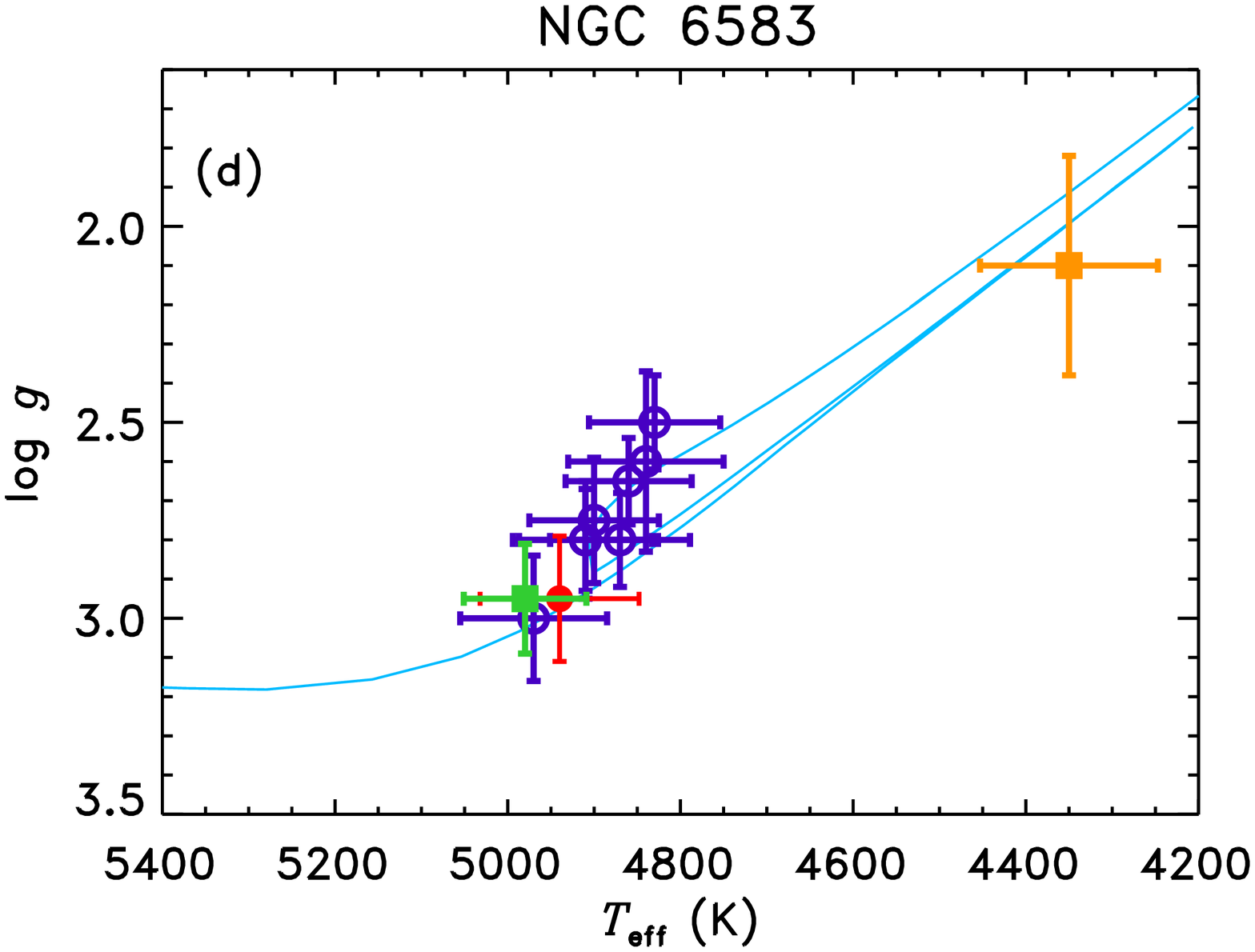}
\caption{ \teff--$\log g$ diagrams of the four open clusters in this study together with their respective isochrones.  The symbols are the same as in Figure~\ref{fig:CMDs}.  \label{fig:CMDs2}}
\end{figure*}

In Figure~\ref{fig:CMDs2}, we find overall good agreement between our derived \teff\ and $\log g$ and the values predicted from  isochrones.  However, the classification of objects as either candidate RC stars (circles) or candidate non-RC stars (squares) from the photometry (Figure~\ref{fig:CMDs}) do not necessarily persist when plotted in this phase space.  Our RC stars tend to scatter along a diagonal line in \teff--$\log g$  but at a given temperature, the clump and first ascent stars appear to cleanly separate in $\log g$, as is especially apparent in NGC~2204.  In that cluster, two of the RC candidate stars appear to be first ascent stars just below the luminosity bump.   The same is true of one of the RC candidates in Collinder~110. These stars have revised classifications of `RGB' in Tables \ref{tab:oc_starlist1} and \ref{tab:oc_starlist2}. All of the clump candidates in NGC~2506 appear to be bona fide RC stars. The picture in NGC~6583 is still murky. The two hottest RC candidates may be first ascent stars, and their revised classifications in Table \ref{tab:oc_starlist4} is `RGB?.' In the discussion that follows,  `candidate RC' stars refers to the stars that have initial classifications of `RC;'  their revised classification may be different.

Comparing Figures \ref{fig:CMDs} and \ref{fig:CMDs2}, we were surprised to find that the single non-RC star that we observed in NGC~2506 
has the hottest spectroscopic temperature (5250~K), even though it has the reddest $B-V$ of the stars we observed. 
Our spectroscopic measurements place NGC~2506~3265 somewhat off of the isochrone, but we note that this star has the lowest signal-to-noise ratio (S/N) in our entire sample (only 68).   
 Using the color--temperature relations of \cite{2005ApJ...626..465R}, the $(B-V)_0$ color yields $T_{\rm eff}=4690$~K.
To test the source of this discrepancy,  we computed color--temperatures from two independent photometric sources  in the literature. Optical $(V-I)_0$ photometry from \cite{2012MNRAS.425.1567L} yields a hotter temperature of $T_{\rm eff}=4820$~K, while 2MASS $(J-K)_0$ (using \citealt{2000AJ....119.1448H} color transformation) yields an even higher $T_{\rm eff}=5050$~K. This latter value is most consistent with the spectroscopic temperature but is still 200~K cooler. All of the literature photometry of  NGC~2506~3265 is consistent with the RC and bump magnitude of the isochrones. The optical colors correspond to the luminosity bump, while the infrared color prefers the RC. Thus, this star's evolutionary stage remains ambiguous.
It is also  possible that the star is a non-member considering that both its \vhelio\ and [Fe/H] appear to be deviant, although neither parameter is sufficiently deviant compared to the errors to confidently confirm its non-membership. 
Comparing our \vhelio\ measurement ($80.3\pm0.3$~\kms) to the literature suggests that the star may in fact be a binary. \cite{2014AJ....147..138C} finds $83.7\pm1.4$ while \cite{2007AA...470..919M} find $85.33\pm0.46$~\kms.

\subsubsection{[Fe/H] and Equivalent Width Comparison}
\begin{figure*}[tb]
\includegraphics[width=0.5\textwidth]{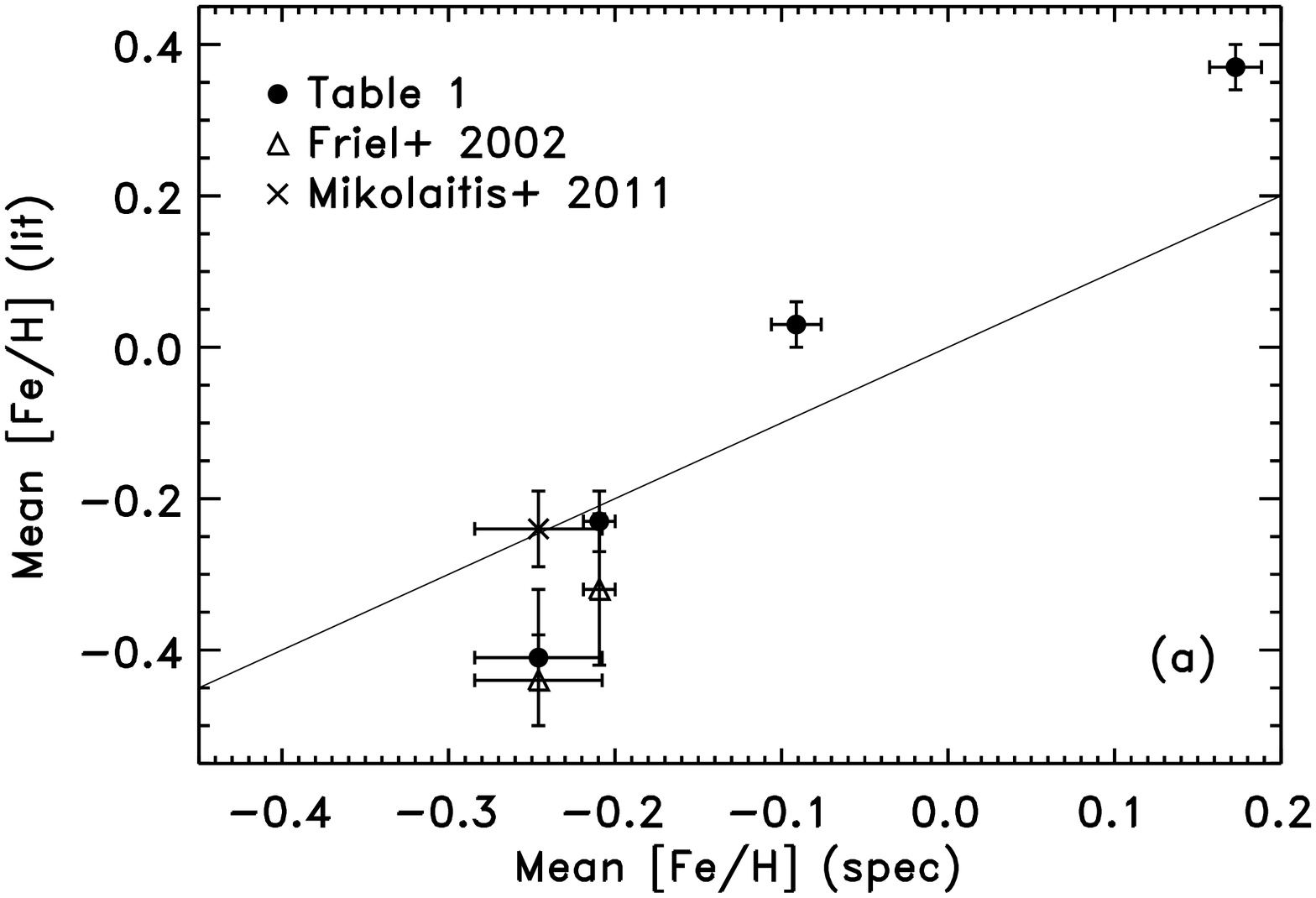}
\includegraphics[width=0.5\textwidth]{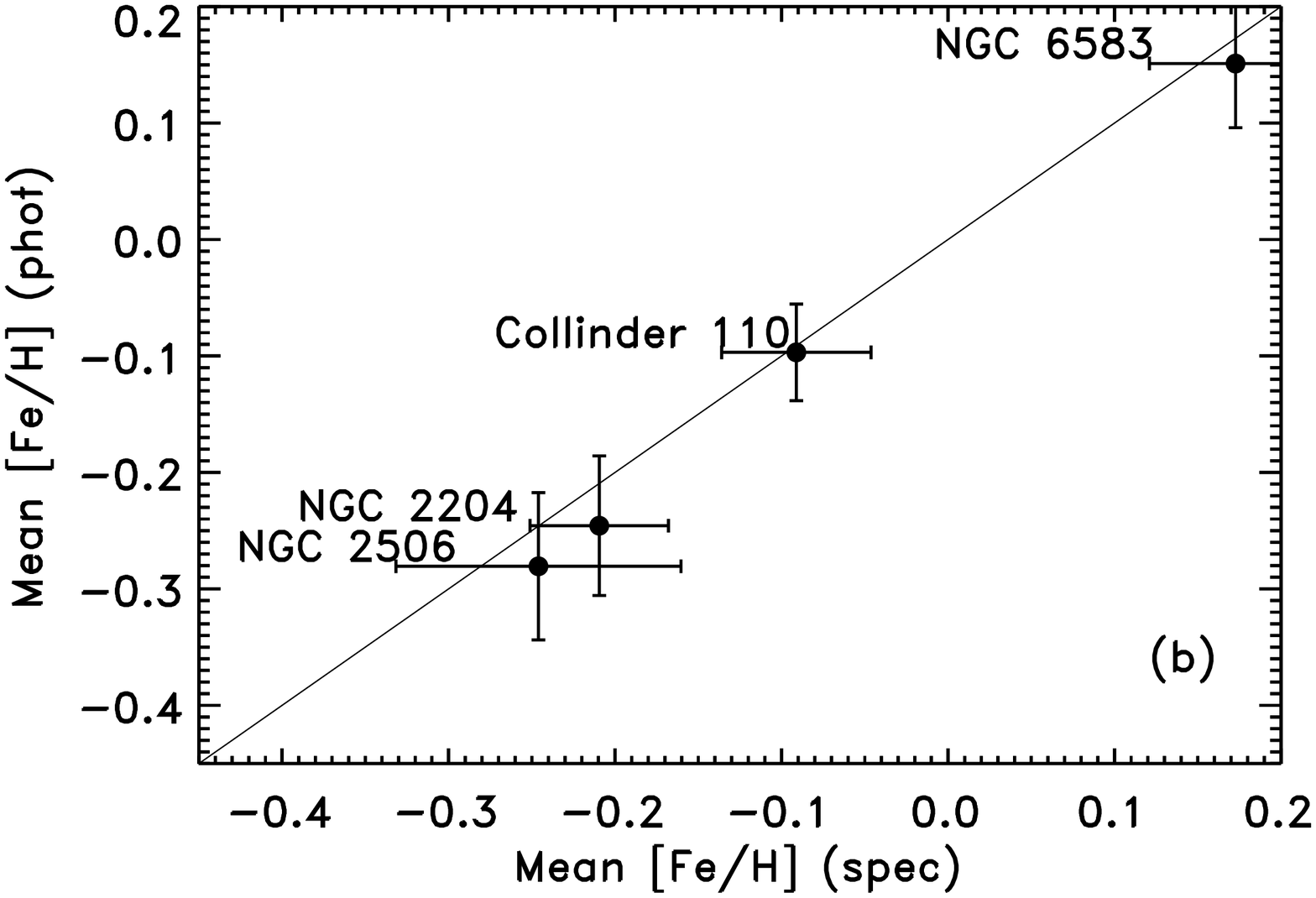}
\caption{Comparison between our average cluster [Fe/H]  and measurements from the literature  (a). Comparison between the average [Fe/H] measurements from the full spectroscopic determination and from one that uses \teff\ and $\log g$ from the photometry  (b).
\label{fig:compfeh}}
\end{figure*}

In  Table~\ref{tab:oc_rv} and Figure~\ref{fig:compfeh}, we compare our average cluster metallicities to those in the literature. Most of the literature sources are those listed in Table~\ref{tab:clusters}.
For the two most metal-poor clusters, we get excellent agreement with at least one literature source, although other literature sources disagree outside the uncertainties. For the more metal-rich clusters, we find lower metallicities than the literature studies do. To test whether the cause of this discrepancy arises from our stellar parameter derivation, we also compute the metallicities using the same EqW measurements but using \teff\  derived from the dereddened colors and magnitudes of the stars and  $\log g$ derived from the stellar masses estimated from the isochrones. The bottom panel of Figure~\ref{fig:compfeh} shows that comparison. The photometric cluster metallicities are in good agreement with the spectroscopic ones but are systematically lower.

\begin{figure*}[tb]
\includegraphics[width=0.5\textwidth]{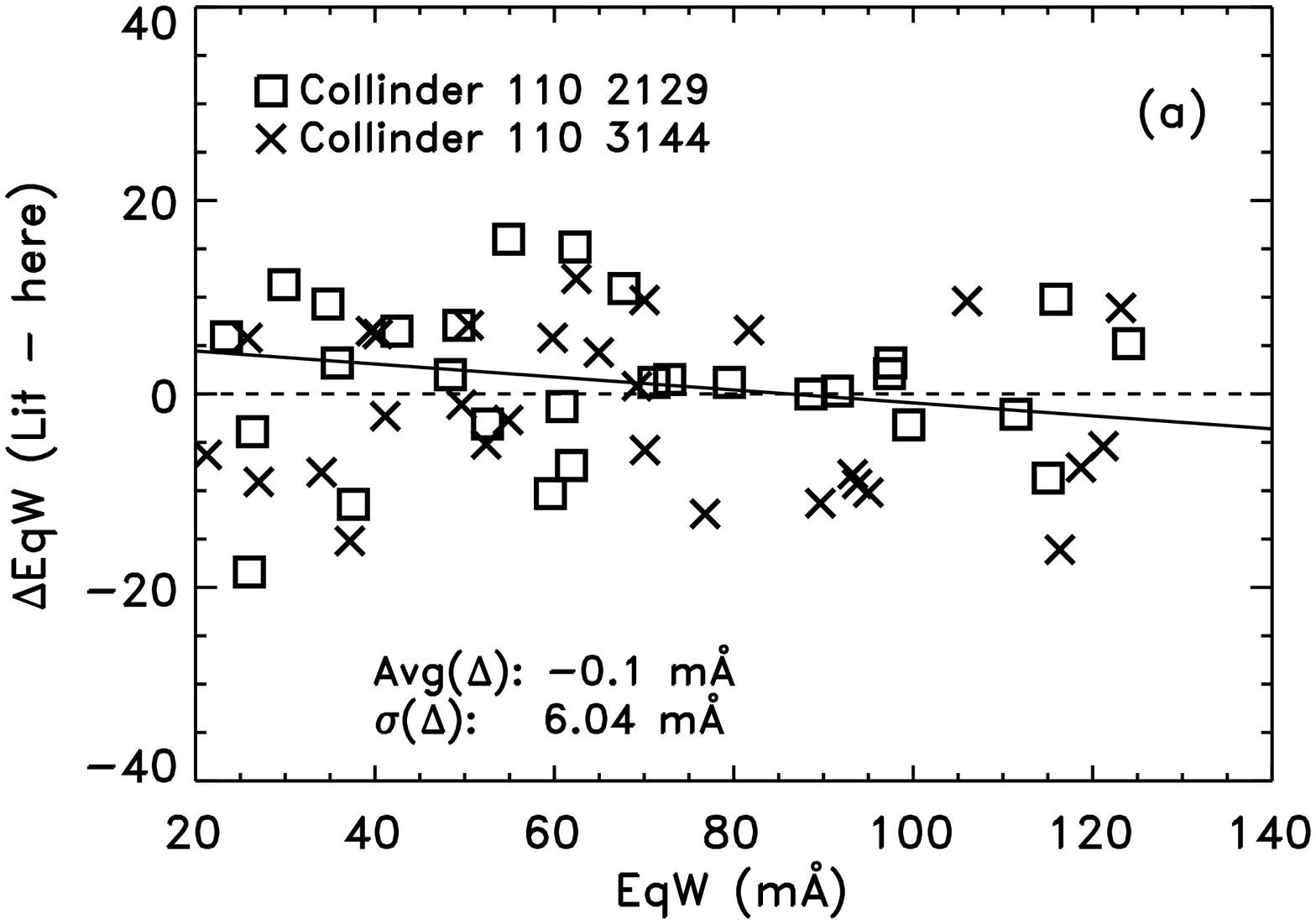}
\includegraphics[width=0.5\textwidth]{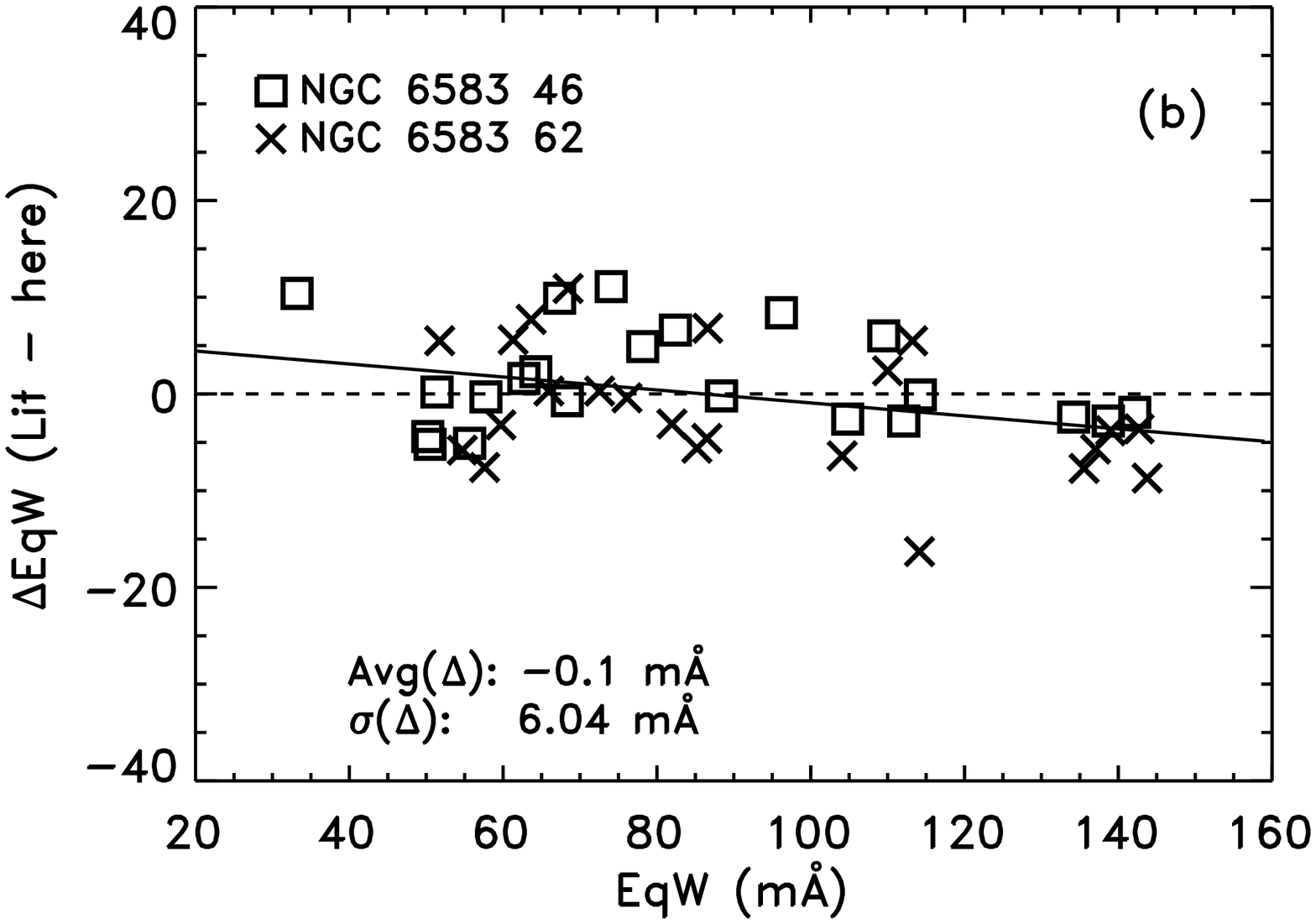}
\caption{Comparison of  our iron line equivalent width measurements  to literature measurements for stars in Collinder 110 \citep[a,][]{2010AA...511A..56P} and NGC~6583 \citep[b,][]{2010AA...523A..11M}. 
The solid lines show  linear fits to the data, while the dashed lines denote the ideal case of $\Delta=0$.  The mean and standard deviation of $\Delta$EqW is also shown on the plot.
\label{fig:compeqw}}
\end{figure*}

Another possible cause of the discrepancy may be in the EqW measurements.  The presence of many, weak unresolved metal lines could lead to systematic differences in the continuum placement.  Fortuitously, there are two stars each in Collinder~110 and  NGC~6583 with literature EqW abundances.
\cite{2010AA...511A..56P} measured abundances of Collinder~110  stars 2129 and 3144, and there are 29 iron lines overlapping our studies.  Similarly, 
 \cite{2010AA...523A..11M} measured abundances  of the NGC~6583 stars 46 and 62 with 22 lines overlapping our studies. 
 Figure~\ref{fig:compeqw} shows that our measurements agree quite well, and there is no systematic underestimation of EqWs in this study.
The  mean and standard deviation of $\Delta$EqW are similar in size to that attained by \cite{2010AA...511A..56P}, when they compared their results to earlier studies.

\subsection{Rotational Velocities}
\label{sec:rot}
The projected rotational velocities (\vsini) were measured by fitting the profiles of up to six relatively isolated Fe lines in our line list (5307.36, 5638.26, 5855.08, 6151.62, 6165.36,  and 6750.15~\AA).   In addition to the rotation, there are two other major sources of broadening that must be accounted for: the line spread function (or instrumental broadening) and the macroturbulent broadening ($\zeta$). The former can be measured from the Th--Ar  lines appearing in the same echelle order as the Fe lines. 
Discriminating uniquely between \vsini\ from $\zeta$  is difficult (if not impossible) at the spectral resolution and S/N of our data.  Instead, we estimate $\zeta$ using previously defined relationships between $\zeta$ and other standard stellar parameters. We tested two different prescriptions: the $\zeta$(\teff, LC) relationships of \cite{2007A&A...475.1003H}, where `LC' refers to the luminosity class, and the $\zeta$($\log$ \teff, $\log L/L_\sun$) relationships of \cite{2008AJ....135..209M}.
The former prescription tended to give larger $\zeta$ by $\sim$0.5--2.5~\kms\  compared to the latter.  We adopted the \cite{2008AJ....135..209M} prescription since we have fairly accurate luminosities for our stars.  We then generated synthetic spectra of varying \vsini\ for each of the six spectra lines, and used a $\chi^2$ minimization to select the best \vsini.

The results of the broadening analysis (\vsini, $\sigma_{v \sin i}$, and $\zeta$) are given in   columns 11--13 of Table~\ref{tab:stell_param}. 
The majority of the RGs in this study are very slow rotators, where the instrumental and macroturbulent broadening fully account for the line profile, and in some cases over accounts for it.  Upper limits are provided for these stars, which are indicated with $\sigma_{v \sin i}=0.0$~\kms.
Only three stars have \vsini\ that are not upper limits.
The star with the largest rotation, Collinder~110~2119, has \vsini$=5.9\pm0.3$~\kms. Its \vsini\ was also measured by \cite{2014AJ....147..138C} via cross-correlation using high-resolution,  low S/N data and was found to be $6.3\pm0.8$~\kms, consistent with this result.  The remaining Collinder 110 stars overlapping that study were all upper limits consistent with what is found here.

\subsection{Lithium}
\label{sec:lithium}

\ali\ were measured by fitting spectra synthesized with MOOG to the observed data between 6706.7 and 6708.4~\AA\ using the line list from \cite{2009ApJ...698..451G}.  The spectra were fit by hand, making small adjustments to the overall continuum level, velocity scale, and broadening to get a good match between the observed and synthetic spectra. The \ion{Li}{1} resonance lines are blended with a \ion{Fe}{1} line and CN lines at the resolution of our spectra.  We made adjustments to the abundances of these elements first before adjusting \ali. Our synthesis fits assume local thermodynamic equilibrium (LTE), and we correct for  NLTE effects by interpolating the grid of  \cite{2009A&A...503..541L} corrections to each  stars' observed parameters.

The results of the synthesis fitting are given in Table~\ref{tab:stell_abun}.  The EqW of the Li feature corresponding to the synthesis measurement or limit is also provided in the table.
For some spectra, the ``best fit'' was still not a great match to the data, and we therefore provide a reliability parameter to flag the cases where the fits were more uncertain.  As an example, NGC~2506 is the most metal-poor cluster in the sample and, because of observing conditions,  the average S/N achieved for these stars is lower than all of the other clusters. Therefore, the spectra had both the weakest features to fit and the largest noise. The stars with a quality  parameter of `A' are good fits, whereas `B' indicates more uncertain fits.

\subsection{\cratio}
The \cratio\ ratio was measured  by fitting the spectral region between 8001 and 8006~\AA, containing four $^{12}$CN features (one blended with an \ion{Fe}{1} line) and one $^{13}$CN feature. We used the line list from \cite{carlberg12} to generate synthetic spectra. A linear correction to the continuum is fit, and the observed spectrum is cross-correlated with the synthetic spectrum to correct for small mismatches between the velocity scales.  We keep the ratio of C/N fixed at 1.5 but allow the total abundance of the two elements to vary to fit the 
 $^{12}$CN features. Then we vary the \cratio\ to find the best fit to the $^{13}$CN features. Telluric features are abundant in this part of the spectrum. In most cases the telluric lines are weak, and we account for their presence by adding in their contribution using the \cite{hinkle00} atlas telluric spectrum, broadened to our instrumental resolution and scaled to fit the depth of the features.   
 
 There is one strong telluric feature near 8007.5~\AA\ that caused significant problems.  NGC~2204 was observed at an unfortunate  RV that placed the $^{13}$CN lines at nearly the same wavelength as this strong feature.   For NGC~2506, the blue wing of this strong telluric feature is blended with the $^{13}$CN line. For the stars in these two clusters, we corrected the spectra using the telluric standard stars that were observed on each night.  These standards were observed at  airmasses that bracketed the airmass of the cluster stars.   We divided each of the cluster  stellar spectra with the hot star spectrum that was closest in airmass, adjusting the velocity and overall scaling of the hot star spectrum to get the cleanest division.
  In seven cases, the  $^{13}$CN feature  was not recoverable. 
 For the NGC~2204 stars where  the feature was recovered, it should be noted that the telluric feature was always  stronger that the underlying stellar feature, and uncertainties in the telluric division may be large. Therefore, all of the \cratio\ measurements for NGC~2204 should be used with caution. 
 The results of the \cratio\ analysis are given in Table~\ref{tab:stell_abun}, which includes a quality parameter. Stars with measurements have a quality parameter ranging from `A' through `D,' which corresponds to decreasing quality of the fits. Those with quality of `X' indicates that \cratio\ was not measurable.  The two lowest quality measurements (`D' and `X') apply only to the stars that had the hot star correction to the telluric features.

\begin{deluxetable*}{lrrrrrrrrrr}
\tablecolumns{10}
\tablewidth{0pc}
\tabletypesize{\scriptsize}
\tablecaption{Stellar Abundances \label{tab:stell_abun}}
\tablehead{
   \colhead{Cluster} &
   \colhead{Star} &
   \colhead{EqW} &
   \colhead{\ali} &
   \colhead{\ali$_{\rm NLTE}$} &
   \colhead{$\sigma_{A({\rm Li})}$\tablenotemark{a}} &
   \colhead{Quality} &
   \colhead{\cratio} &
   \colhead{$\sigma_{^{12}{\rm C}/^{13}{\rm C}}$\tablenotemark{b}} &
   \colhead{Quality} \\
   \colhead{} &
   \colhead{} &
   \colhead{(m\AA)} &
   \colhead{(dex)} &
   \colhead{(dex)} &
   \colhead{(dex)} &
   \colhead{} &
   \colhead{} &
   \colhead{} &
   \colhead{} &
    \colhead{}  }
\startdata
Collinder 110  &  1134  &  $  8.7$  &  $ 0.62$  &  $ 0.79$  &  0.15  &  A  &  17  &   5  &  A \\
Collinder 110  &  2119  &  $  9.7$  &  $ 0.68$  &  $ 0.84$  &  0.16  &  A  &  22  &   5  &  B \\
Collinder 110  &  2129  &  $  4.8$  &  $ 0.34$  &  $ 0.51$  &  0.00  &  A  &  12  &   5  &  A \\
Collinder 110  &  2223  &  $  2.9$  &  $ 0.18$  &  $ 0.34$  &  0.00  &  A  &  15  &   5  &  C \\
Collinder 110  &  3122  &  $ 63.7$  &  $ 1.34$  &  $ 1.54$  &  0.22  &  A  &  18  &   5  &  B \\
Collinder 110  &  3144  &  $ 13.0$  &  $ 0.68$  &  $ 0.88$  &  0.00  &  B  &  10  &   5  &  C \\
Collinder 110  &  3244  &  $  6.9$  &  $ 0.52$  &  $ 0.69$  &  0.15  &  A  &  16  &   5  &  B \\
Collinder 110  &  4260  &  $  4.0$  &  $ 0.30$  &  $ 0.47$  &  0.00  &  A  &  17  &   5  &  B \\
Collinder 110  &  5125  &  $  3.8$  &  $ 0.45$  &  $ 0.58$  &  0.00  &  A  &  20  &   5  &  B \\
NGC 2204  &  1124  &  $  3.7$  &  $ 0.31$  &  $ 0.46$  &  0.00  &  B  &  \nodata  &  \nodata  &  X \\
NGC 2204  &  1212  &  $  5.9$  &  $ 0.59$  &  $ 0.73$  &  0.13  &  A  &  \nodata  &  \nodata  &  X \\
NGC 2204  &  1330  &  $  1.7$  &  $ 0.07$  &  $ 0.20$  &  0.00  &  A  &  16  &   5  &  D \\
NGC 2204  &  2212  &  $  0.3$  &  $-1.22$  &  $-0.95$  &  0.00  &  A  &  12  &   5  &  D \\
NGC 2204  &  2229  &  $  4.8$  &  $ 0.47$  &  $ 0.62$  &  0.00  &  B  &  12  &   5  &  D \\
NGC 2204  &  2311  &  $  3.6$  &  $ 0.19$  &  $ 0.36$  &  0.00  &  A  &  12  &   5  &  C \\
NGC 2204  &  2330  &  $ 35.6$  &  $ 1.40$  &  $ 1.52$  &  0.14  &  A  &  12  &   5  &  C \\
NGC 2204  &  3205  &  $  3.9$  &  $ 0.39$  &  $ 0.53$  &  0.17  &  A  &  \nodata  &  \nodata  &  X \\
NGC 2204  &  3215  &  $  7.4$  &  $ 0.59$  &  $ 0.75$  &  0.15  &  B  &   8  &   5  &  D \\
NGC 2204  &  3321  &  $  8.3$  &  $ 0.75$  &  $ 0.89$  &  0.13  &  A  &   9  &   5  &  D \\
NGC 2204  &  4115  &  $ 50.8$  &  $ 1.33$  &  $ 1.50$  &  0.16  &  A  &  27  &   5  &  C \\
NGC 2204  &  4116  &  $ 43.7$  &  $ 1.23$  &  $ 1.40$  &  0.16  &  A  &  \nodata  &  \nodata  &  X \\
NGC 2204  &  4119  &  $ 46.7$  &  $ 1.21$  &  $ 1.39$  &  0.19  &  A  &  16  &   5  &  C \\
NGC 2204  &  4211  &  $  5.6$  &  $ 0.49$  &  $ 0.64$  &  0.16  &  A  &  \nodata  &  \nodata  &  X \\
NGC 2204  &  4223  &  $  1.1$  &  $ 0.10$  &  $ 0.21$  &  0.00  &  A  &  15  &   5  &  D \\
NGC 2204  &  4303  &  $  2.2$  &  $ 0.09$  &  $ 0.24$  &  0.00  &  B  &  10  &   5  &  D \\
NGC 2204  &  5352  &  $ 17.9$  &  $ 1.00$  &  $ 1.15$  &  0.17  &  B  &  \nodata  &  \nodata  &  X \\
NGC 2204  &  5980  &  $  2.1$  &  $ 0.12$  &  $ 0.27$  &  0.00  &  B  &  18  &   0  &  D \\
NGC 2204  &  6330  &  $  7.9$  &  $ 0.62$  &  $ 0.77$  &  0.15  &  B  &  13  &   5  &  D \\
NGC 2506  &  2380  &  $  5.6$  &  $ 0.43$  &  $ 0.59$  &  0.00  &  B  &  \nodata  &  \nodata  &  X \\
NGC 2506  &  3265  &  $  9.8$  &  $ 1.00$  &  $ 1.11$  &  0.00  &  B  &  21  &   5  &  B \\
NGC 2506  &  4138  &  $ 10.1$  &  $ 0.76$  &  $ 0.90$  &  0.15  &  B  &  10  &   0  &  C \\
NGC 2506  &  4205  &  $  5.4$  &  $ 0.50$  &  $ 0.64$  &  0.14  &  B  &  10  &   5  &  D \\
NGC 2506  &  4240  &  $  5.8$  &  $ 0.48$  &  $ 0.63$  &  0.00  &  B  &   9  &   5  &  C \\
NGC 6583  &  10  &  $262.3$  &  $ 1.49$  &  $ 1.74$  &  0.33  &  A  &  24  &   5  &  A \\
NGC 6583  &  33  &  $ 20.2$  &  $ 0.88$  &  $ 1.07$  &  0.22  &  A  &  21  &   5  &  A \\
NGC 6583  &  34  &  $ 21.3$  &  $ 0.89$  &  $ 1.09$  &  0.20  &  B  &  23  &   5  &  A \\
NGC 6583  &  38  &  $ 31.9$  &  $ 1.13$  &  $ 1.32$  &  0.20  &  A  &  22  &   5  &  A \\
NGC 6583  &  39  &  $  2.6$  &  $-0.09$  &  $ 0.13$  &  0.00  &  A  &  21  &   5  &  A \\
NGC 6583  &  42  &  $ 22.1$  &  $ 0.88$  &  $ 1.09$  &  0.00  &  B  &  19  &   5  &  A \\
NGC 6583  &  46  &  $ 13.8$  &  $ 0.76$  &  $ 0.94$  &  0.20  &  A  &  23  &   5  &  B \\
NGC 6583  &  50  &  $ 22.5$  &  $ 1.06$  &  $ 1.23$  &  0.22  &  A  &  22  &   5  &  A \\
NGC 6583  &  62  &  $ 21.5$  &  $ 0.96$  &  $ 1.14$  &  0.22  &  A  &  23  &   5  &  A \\
NGC 6583  &  72  &  $ 61.4$  &  $ 1.50$  &  $ 1.67$  &  0.24  &  A  &  31  &   5  &  C \\
NGC 6583  &  92  &  $ 41.3$  &  $ 1.36$  &  $ 1.52$  &  0.18  &  A  &  27  &   5  &  A \\
\enddata
\tablenotetext{a}{Values of 0.0 refer to upper limits.}
\tablenotetext{b}{Values of 0.0 refer to lower limits.}
\end{deluxetable*}

\section{Discussion} 
\label{sec:discuss}
   
\subsection{Lithium Distribution}
\begin{figure}[tb]
\includegraphics[width=0.5\textwidth]{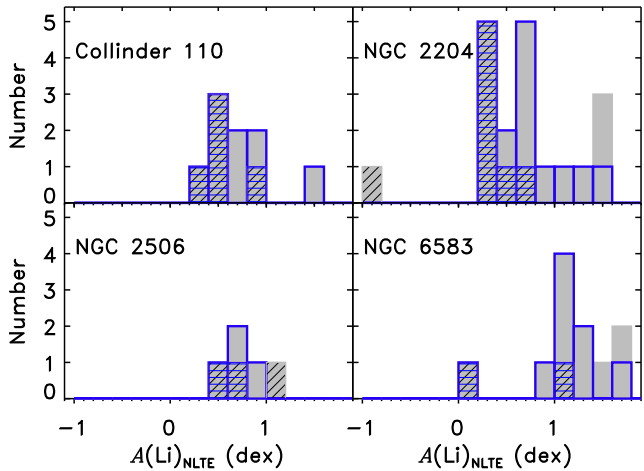}
\caption{The distribution of the non-LTE  \ali\  for all stars observed in each cluster
 (shaded gray histograms)  and the subset of candidate RC stars (open blue histograms).   The hatched regions denote  stars with only upper limits on \ali\ in the full sample (diagonal lines) and the RC sample (horizontal lines).  \label{fig:li_histo}}
\end{figure}

In Figure~\ref{fig:li_histo}, we show the distribution of \ali$_{\rm NLTE}$ for the four clusters in this study, with histograms for both the full observed sample and the subset of RC candidate stars.  
The non-RC stars tend to fall at either extremes of the distributions, but are more commonly at the high extreme.
For three of the clusters, the majority of the RC candidates  fall into a single distribution that peaks around $\sim0.7$~dex and is composed of mostly upper limit measurements.  Two of the those clusters (Collinder~110 and NGC~2204)  have RC candidates in a second group or tail that extends into the Li-rich regime.
In contrast, the fourth cluster (NGC~6583, both  the most-metal rich and youngest cluster in this study) shows a peak in the distribution at much higher \ali\ with very few upper limit measurements. Only a single  RC star has significantly lower \ali.

 These bimodal-like lithium distributions are similar to  three clusters with \ali\ dichotomies reported previously in the literature for NGC~752, NGC~3680, and IC~4651 (\citealt{pilachowski88}, \citealt{2001A&A...374.1017P}, \citealt{2004A&A...424..951P}).  
All three clusters have similar ages (1.5--1.9~Gyr) and are older than two of our clusters (NGC~6583 and Collinder~110) but younger than the other two (NGC~2204 and NGC~2506).
The dichotomy of \ali\ in these three literature clusters has been interpreted as discriminating between first ascent RGs and RC stars in each cluster, with the higher \ali\ corresponding to the former, and lower \ali\ to the latter.  This interpretation is also borne out in Collinder~110 and NGC~2204, where it is clear from  Figure~\ref{fig:CMDs2} that the highest Li stars (filled symbols) are the ones that appear to be first ascent RGB stars and not RC stars.

\begin{figure*}[tb]
\includegraphics[width=0.5\textwidth]{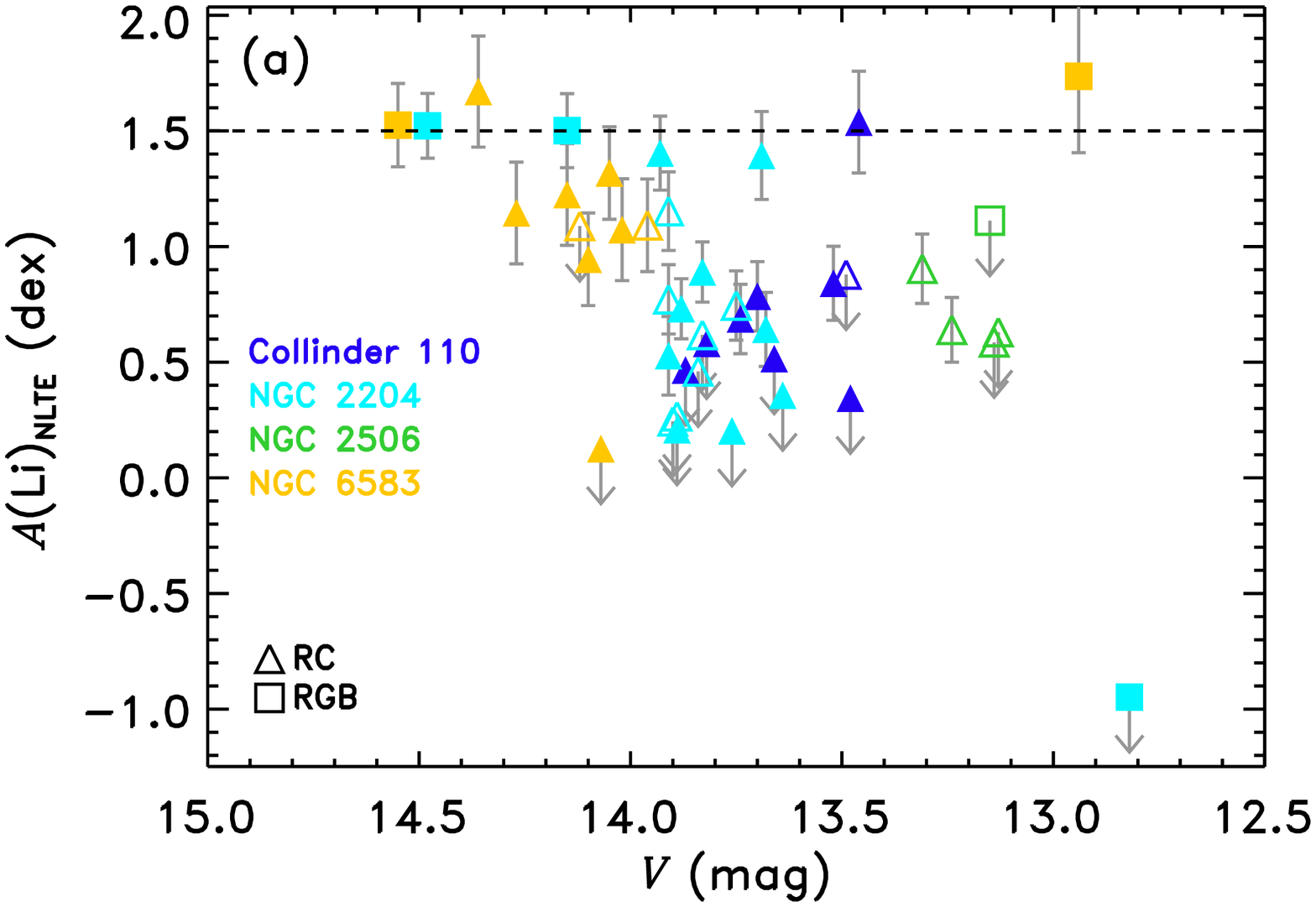}
\includegraphics[width=0.5\textwidth]{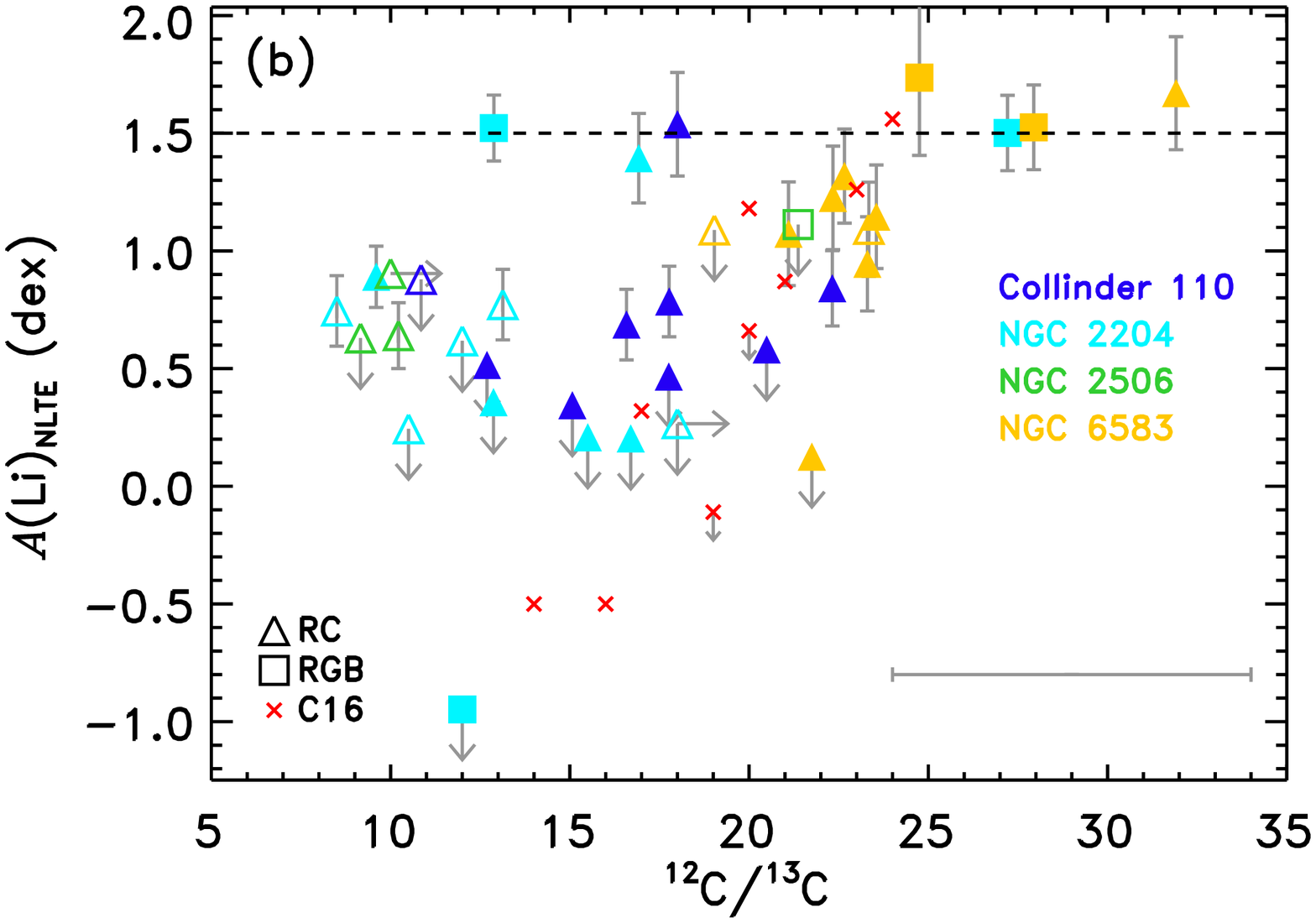}
\caption{\ali$_{\rm NLTE}$ as a function of $V$ magnitude (a), and \cratio\  (b). Triangles are RC candidates, while squares represent likely RGB stars. Arrows indicate limits. Filled symbols are the quality `A' lithium measurements (most reliable), while open symbols are quality `B.' In the second plot, the $\times$'s show the results from \cite{2016ApJ...818...25C}, and a representative error bar for \cratio\ is shown in the lower right. \label{fig:li_results}}
\end{figure*}

However, because RC should be more common than first ascent RGs, one would expect all of the clusters to have more low \ali\ stars than high \ali\ stars.  This is not the case for NGC~3680 (or NGC~6583 in this work), which led \cite{2001A&A...374.1017P} to suggest that the high \ali\ stars in NGC~3680 were in fact the RC stars.   An expanded sample of giants in NGC~3680 by \cite{2009AJ....138.1171A} confirmed this Li dichotomy.
 NGC~3680 is the oldest and most metal-poor ([Fe/H]$\sim-0.14$~dex) of those three literature clusters, but NGC~6583 is the youngest and most metal-rich of the clusters studied here. 
Since we have previously estimated evolutionary stages for our cluster stars, we can explore the Li distribution of each cluster in more detail. In Figure~\ref{fig:li_results}, we show  \ali\ as a function of $V$ magnitude and \cratio.  In general, the  RGB stars (identified as squares) show the largest abundances, while the candidate RC stars (triangles) show depleted abundances. This is true even in NGC~6583;  the RC stars have depleted \ali\ compared to the first ascent RGB stars, but the depletion in this cluster is less than that in the other three clusters. The apparent dichotomous \ali\ distribution in this cluster is in fact three different levels of \ali: RGB stars with the least Li dilution, RC stars with moderate Li dilution, and one RC star with significant dilution. 

Our four cluster sample suggests that if  Li-replenishment  is commonly occurring at the He flash, the subsequent destruction of Li reduces the surface abundance below the RGB levels.
 A few outliers are worth noting. First, in NGC~6583 the most luminous star has \ali\ comparable to the other RGB star, whereas the most luminous star in NGC~2204 has severely depleted levels.  This implies that the NGC~2204 luminous star is on the AGB, not the RGB. In all of the clusters except NGC~2506, there are one or more candidate RC stars with \ali\ more similar to the RGB star levels. RC stars are plotted as red circles in Figures~\ref{fig:CMDs} (which was used to determine their RC candidacy) and \ref{fig:CMDs2}.   We find that all but one of the candidate RC stars with high Li are closer matches to the first ascent RGB in Figure~\ref{fig:CMDs2} and were reclassified based on those positions. The exception is NGC~6583~72, whose classification is ambiguous  in Figure~\ref{fig:CMDs2}.

The second panel in Figure~\ref{fig:li_results} shows \ali\ as a function of \cratio. Most of the stars fall along a trend of decreasing \cratio\ with decreasing \ali. This behavior is expected since both quantities are altered by mixing in stars. The stars along the linear trend follow the same distribution as the open cluster stars studied by \cite{2016ApJ...818...25C}.  There are again some exceptions.  There is a grouping of eleven stars with low \cratio\ but \ali$\sim 0.6$~dex. Seven of these have limits in one of the parameters such that they could  fall along the trend.  Four stars, on the other hand, are still outliers. All of the stars are in the clusters that had the difficult telluric removal, so it is possible that the low \cratio\ measurements are spurious. However,  our confidence in the validity of the measurements are due to the facts that (1) the $^{13}$CN feature is stronger (more easily detectable) for low \cratio, and (2) owing to the  blend of lines creating the $^{13}$CN feature, the width of the line would be difficult to mimic with residuals from the telluric removal.

\subsection{Test of the He Flash Li Enrichment Mechanism}
This sample of stars was selected specifically to test the hypothesis that Li is generated at the He flash.  In the \citetalias{2011ApJ...730L..12K} sample, all of the Li-rich stars that have \cratio\ available have \ali$>1.98$~dex.   This minimum \ali\ in the \citetalias{2011ApJ...730L..12K}  sample is influenced by two factors. First, the newly discovered RGs in \citetalias{2011ApJ...730L..12K} were pre-screened for the presence of the Li line in low resolution spectra, leading to a temperature-dependent minimum \ali. Second, the lowest \ali\ stars in both the new RG and literature RG samples in \citetalias{2011ApJ...730L..12K} do not have \cratio\ measurements. 
None of our stars have Li abundances at this level, implying a low occurrence rate for Li-rich stars. Using the binomial statistics   in \cite{2003ApJ...586..512B}, we can estimate the $1\sigma$ upper bound
for the occurrence rate of RC clump stars more Li-rich than $1.98$~dex to be $<5$\%.  This result is the same if we use all 38 photometrically identified RC candidates or if we reduce the sample size to 35 to account for the ones that may be RGB stars.
The lack of a Li-rich star in our sample is still consistent with the \citetalias{2011ApJ...730L..12K} field giant study, which estimated the incidence to be around 1\%.  

\begin{figure}[tb]
\includegraphics[width=0.5\textwidth]{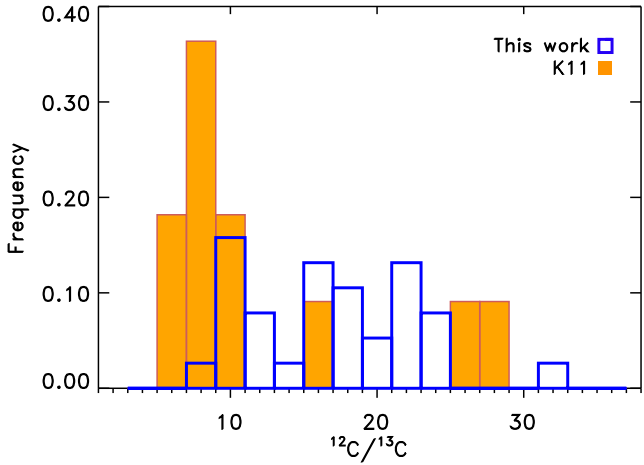}
\caption{Distribution of \cratio\ in the 38 RC candidates in this work (open histogram) and for the Li-rich stars with $\log L/L_\sun < 2.5$ and \teff$>4750$~K in \citetalias{2011ApJ...730L..12K} (filled histogram). The temperature and luminosity limits were imposed on the \citetalias{2011ApJ...730L..12K} sample to restrict it to stars most similar to the RC stars in this work.
\label{fig:cratio_hist}}
\end{figure}

If we factor  \cratio\ into our consideration, we can perform a more stringent test of the fraction of stars that go through a Li-enriched phase early in the core He-burning stage.
In Figure~\ref{fig:cratio_hist}, we show the \cratio\ distribution of the subset of the \citetalias{2011ApJ...730L..12K} sample that overlaps our RC sample in parameter space. We remove stars from the \citetalias{2011ApJ...730L..12K} paper that have $\log L/L_\sun > 2.5$ and    \teff$<4750$~K. The first cut removes four luminous RGs in \citetalias{2011ApJ...730L..12K} that are outside of the mass range and are near the second dredge-up phase on the early-AGB (see \citealt{2000A&A...359..563C}).  The second cut removes cooler RGs that are more likely to be at the luminosity bump. Overlaid in Figure~\ref{fig:cratio_hist} is the \cratio\ distribution of our sample of 38 RC candidates. The distributions are very different. The majority of the Li-rich \citetalias{2011ApJ...730L..12K} stars have \cratio$\lesssim10$.  Such a low \cratio\ is a signature of deep mixing that dredges up material from the vicinity of the H-burning shell, where \cratio\ is 3.5 \citep{1965ApJ...141..688C}.
This implies that the mechanism responsible for the enhanced Li requires (or is accompanied by) very deep mixing.  Because \cratio\ of the stellar atmosphere can only be lowered by mixing, this signature of low \cratio\ will persist even when the lithium levels  return to normal low values. Therefore, the \cratio\ distribution of our RC sample is inconsistent with the hypothesis that most/all stars of 1.5--2.2~\msun\ experience a brief Li-rich stage. Instead, the Li-rich stage must only be experienced by a small fraction of stars.

We can identify the subset of stars in our sample that likely went through a Li-rich phase as the outlier group of low \cratio\  stars we identified in Figure~\ref{fig:li_results}.  
If the four stars that are clear outliers (i.e., no limit on either abundance measurement) are the only true outliers,  then only $13.8\%^{+8.7\%}_{-4.1\%}$ of the RC stars went through a Li-rich phase.   The fraction is larger if we include all 11 stars in the low \cratio\ outlier group, which results in an occurrence rate of $37.9\%^{+9.5\%}_{-7.9\%}$. The RC sample for these two calculations  only use the RC candidates without strong evidence for reclassification (i.e., the `reclassified' RC candidates in Collinder~110 and NGC~2204 are removed) and only considers stars with \cratio\ measurements. These constraints yield 29 RC stars.

 \subsection{Confounding Factors: Stellar Rotation and Other Physics}
Our interpretations may also be affected by the unknown rotational histories of the individual stars. The main sequence progenitors of the stars in the mass range we are studying have a large distribution of rotational velocities, with a typical value of 150~\kms\ \citep{2007A&A...463..671R}.  Despite this large variation on the MS, \cite{2014AJ....147..138C} showed that the red giant descendants of these stars are still almost uniformly slow rotators, a result that is reproduced here (Section \ref{sec:rot}).  This fact makes it difficult to discern which stars were faster or slower MS rotators.  The initial MS rotation affects the depth of mixing and thus the observed surface \ali\ of RGs \citep[see, e.g.,][]{2012A&A...543A.108L}. Furthermore, rotation extends the main sequence lifetime \citep{2010A&A...509A..72E}.  In an effort to explain the extended MS turn-offs of clusters in the Magellanic Clouds, \cite{2011MNRAS.412L.103G} explored the differences of isochrone morphology  of models including large initial rotation compared to non-rotating models. 
They found that the isochrones made with  fast rotating models were nearly indistinguishable from the slow rotating models. 
 The implication to this work is that some of the RGs may be both more massive than we expect and may have experienced extensive rotational mixing.  These stars would be difficult to distinguish from the stars that conform to the non-rotating model assumptions. 

Among intermediate age Milky Way open clusters, some show extended RC morphologies that are inconsistent with simple single stellar population models \citep{2000A&A...354..892G}. Some of this extension is due to the fact that the ages of these clusters are such that the RC stars span the mass range  that delineates the transition from quiescent to He flash core burning, resulting in a primary and secondary RC \citep{1999MNRAS.308..818G}.  However, this cannot explain all of the spread in the MSTO, nor can it be explained with an age spread  \citep{2000A&A...354..892G}.  The authors note that variations in the mass loss rate on the RGB or in the core overshoot efficiencies on the MS (which affects the mass transition for the different types of He core burning onset) could be responsible. If these factors are at play in the clusters of this work, they may also influence the variation of \ali\ observed in our clusters.

\section{Conclusions}
\label{sec:end}
We have measured \ali\ and \cratio\ in a sample of RGs in four Southern open clusters. Most of the stars are RC  stars having masses between  1.6 and 2.2~\msun, which were selected to test the hypothesis that stars in this mass range synthesize Li during the He flash and spend a short fraction of their RC lifetimes as Li-rich stars \citepalias{2011ApJ...730L..12K}. We find seven stars with \ali\ near the threshold of Li-richness at 1.5~dex.  Three of these stars are RGB stars, and all of the four that were identified as candidate RC stars on CMDs  are or could be consistent with the RGB  on spectroscopic \teff$-\log g$ diagrams.

Given the modest sample size, the absence of Li-rich stars in the RC constrains the occurrence (or fractional lifetime) to $<5$\%, which is compatible with the 1\% lifetime quoted by \citetalias{2011ApJ...730L..12K}. However, the majority of the Li-rich RC field giants in \citetalias{2011ApJ...730L..12K} have low \cratio, the result of deep mixing which dredges-up material with \cratio\ near 3.5.  Even if the freshly synthesized Li is destroyed during the RC lifetime, the evidence of the past deep mixing (i.e., low \cratio) should remain.  The RC sample in this study has a \cratio\ distribution that is much higher, demonstrating that these stars experienced less mixing. We therefore conclude that  if a Li-enrichment episode occurs at the He flash, it must only affect a fraction of the stars evolving through that phase.  From a set of outliers in the \ali---\cratio\ distribution, we estimate that fraction to be $13.8\%^{+8.7\%}_{-4.1\%}$. Accounting for  stars with uncertain evolutionary stage and limits in their abundances, we can set a conservative upper limit for the incidence to be  $<47\%$.

\acknowledgments
We are grateful to S.~H. Lee and  H.~B. Ann for providing us their optical photometry of the NGC~2506 red giants in this study. We also thank the referee for valuable feedback that improved this manuscript. 
JKC acknowledges support by an appointment to the NASA Postdoctoral Program at the Goddard Space Flight Center, administered by Universities Space Research Association through a contract with NASA. This paper includes data gathered with the 6.5 m Magellan Telescopes located at Las Campanas Observatory, Chile and  made use of the WEBDA database, operated at the Department of Theoretical Physics and Astrophysics of the Masaryk University.
 
\facility{Magellan:Clay (MIKE)}

\bibliographystyle{aasjournal}

\end{document}